\documentclass[%
 reprint,
nofootinbib,
 amsmath,amssymb,
 aps,
]{revtex4-2}

\usepackage{graphicx}
\usepackage{dcolumn}
\usepackage{bm}
\usepackage{hyperref}
\usepackage[mathlines]{lineno}
\usepackage{multirow}
\usepackage{tabularx}
\usepackage{tabulary}
\usepackage{adjustbox}
\usepackage{xcolor}
\usepackage{tabularray}
\UseTblrLibrary{booktabs}

\renewcommand{\selectlanguage}[1]{}
\newcommand{\rev}[1]{#1}
\newcommand{\revv}[1]{#1}

\begin{document}

\preprint{APS/123-QED}
\title{Improving Undergraduate Astronomy Students' Skills with Research Literature via Accessible Summaries: An \revv{Exploratory} Case Study with Astrobites-based Reading Assignments}

\author{Briley L. Lewis}
\email{brileylewis@g.ucla.edu}
\affiliation{Department of Physics, University of California, Santa Barbara, Santa Barbara, CA 93108 USA}
\affiliation{Department of Physics and Astronomy, University of California, Los Angeles, Los Angeles, CA 90095 USA}

\author{Abygail R. Waggoner}%
\email{awaggoner2@wisc.edu}
\affiliation{University of Virginia, Charlottesville, VA 22904 USA
}%
\affiliation{University of Wisconsin, Madison, WI 53706 USA
}%

\author{Emma Clarke}%
\affiliation{McWilliams Center for Comsology and Department of Physics, Carnegie Mellon University, Pittsburgh, PA 15213 USA
}%

\author{Alison L. Crisp}%
\affiliation{Department of Astronomy, The Ohio State University, Columbus, OH 43210, USA
}%

\author{Mark Dodici}%
\affiliation{David A. Dunlap Department of Astronomy \& Astrophysics, University of Toronto, Toronto, Ontario M5S 3H4 Canada
}%

\author{Graham M. Doskoch}%
\affiliation{Department of Physics and Astronomy, West Virginia University, Morgantown, WV 26506 USA
}%

\author{Michael M. Foley}%
\affiliation{Center for Astrophysics | Harvard \& Smithsonian, Cambridge, MA 0213, USA
}%
\affiliation{Department of Earth and Planetary Sciences, Harvard University, , Cambridge, MA 02138 USA
}%

\author{Ryan Golant}%
\affiliation{Department of Astronomy, Columbia University, New York, NY 10027 USA
}%


\author{Skylar Grayson}%
\affiliation{School of Earth and Space Exploration, Arizona State University, Tempe, AZ 85281 USA
}%

\author{Sahil Hegde}%
\affiliation{Department of Physics and Astronomy, University of California, Los Angeles, Los Angeles, CA 90095 USA
}%


\author{Nathalie Korhonen Cuestas}%
\affiliation{Department of Physics and Astronomy, Northwestern University, 2145 Sheridan Road, Evanston, IL 60208 USA 
}%
\affiliation{Center for Interdisciplinary Exploration and Research in Astrophysics (CIERA), Northwestern University, 1800 Sherman Avenue, Evanston, IL 60201 USA}

\author{Charles J. Law}%
\affiliation{Department of Astronomy, University of Virginia, Charlottesville, VA 22904 USA
}%

\author{R.R. Lefever}%
\affiliation{Zentrum f$\ddot{u}$r Astronomie der Universit$\ddot{a}$t Heidelberg, Astronomisches Rechen-Institut, Heidelberg 69120 Germany
}%


\author{Ishan Mishra}%
\affiliation{Jet Propulsion Laboratory, California Institute of Technology, La Ca$\tilde{n}$ada Flintridge, CA 91011 USA
}%

\author{Mark Popinchalk}%
\affiliation{Department of Astrophysics, American Museum of Natural History, New York, NY 10024 USA
}%

\author{Sabina Sagynbayeva}%
\affiliation{Department of Physics and Astronomy, Stony Brook University, Stony Brook, NY 11794 USA
}%
\affiliation{Center for Computational Astrophysics, Flatiron Institute, New York, NY 10010 USA
}%

\author{Samantha L. Wong}%
\affiliation{Physics Department, McGill University, Montreal, QC H3A 2T8 Canada}%

\author{Wei Yan}%
\affiliation{Department of Mathematics, Computer Science, and Physics, Wartburg College, Waverly, IA 50677 USA
}%

\collaboration{Astrobites Collaboration}

\author{Kaitlyn L. Ingraham Dixie}%
\affiliation{Center for Teaching and Learning, University of Massachusetts Amherst, Amherst, MA, 01003 USA} 
\affiliation{%
Center for Education, Innovation, and Learning in the Sciences (CEILS), University of California, Los Angeles, Los Angeles, CA 90095 USA
}%

\author{K. Supriya}%
\affiliation{%
Center for the Integration of Teaching, Research and Learning (CIRTL)
}%
\affiliation{%
Center for Education, Innovation, and Learning in the Sciences (CEILS), University of California, Los Angeles, Los Angeles, CA 90095 USA
}%

\date{\today}

\begin{abstract}
Undergraduate physics and astronomy students are expected to engage with scientific literature as they begin their research careers, but reading comprehension skills are rarely explicitly taught in major courses. We seek to determine the efficacy of \rev{a reading assignment} designed to improve undergraduate astronomy (or related) majors' perceived ability to engage with research literature by using accessible summaries of current research written by experts in the field. During the 2022-2023 academic year, faculty members from six institutions incorporated \rev{reading assignments} using accessible summaries from Astrobites into their undergraduate astronomy major courses, surveyed their students before and after the activities, and participated in follow-up interviews with our research team. Quantitative and qualitative survey data from 52 students show that \rev{students' perceptions of their abilities with jargon and identifying main takeaways of a paper significantly improved} with use of the tested \rev{assignment template.} Additionally, students \rev{report} increased confidence of their abilities within astronomy after exposure to these \rev{assignments}, and instructors valued a ready-to-use resource to incorporate reading comprehension in their pedagogy. This \revv{exploratory} case study with Astrobites-based \rev{assignments} suggests that incorporating current research in the undergraduate classroom through accessible literature summaries may increase students' confidence and ability to engage with research literature, assisting in their preparation for participation in research careers. 
\end{abstract}

\maketitle

\section{\label{sec:intro}Introduction}

Research experience is an integral part of modern STEM undergraduate education, especially for students who plan to continue on to graduate study \revv{in STEM}, and undergraduates are often highly encouraged to get involved in research early in their careers. Exposure to research during undergraduate education also provides a number of benefits. This includes helping students build research skills, providing a basis for career planning, and encouraging students to develop scientific mindsets \citep{seymour2004establishing}. Exposure to current research in the classroom is also known to have a positive impact on undergraduate learning experiences: engaging with current scientific literature illustrates the open questions and active areas of research in a field and provides real-world motivation for course content \citep{donohue2021integrating,wooten2018investigating}. Recent work exploring Course Based Undergraduate Research Experiences (CUREs) in physics and astronomy further supports the positive benefits of including research in the classroom, including increased persistence in STEM disciplines, development of research skills, stronger identities as scientists, and more \citep{rector2019authentic,werth2022impacts,oliver2023student,werth2023enhancing,hewitt2023development}.

\revv{However, reading} comprehension skills are also rarely explicitly taught in the higher education science classroom, despite the fact that research shows that interventions focused on discipline-based reading and writing skills are beneficial to students. These interventions can empower students to better contextualize their work and engage in scientific discourse, making them more effective, culturally-aware, and well-rounded scientists \citep{duncan2012improving,sorvik2015scientific,szymanski2014instructor,pelger2016popular,lewis2022effects}. These interventions also correlate with success in STEM courses \citep{akbasli2016effect}. The recent Phys21 report from the Joint Task Force on Undergraduate Physics Programs from the American Physical Society, National Science Foundation, and American Association of Physics Teachers further emphasized the need for literature and communication skills as preparation for 21st century careers and the lack of existing preparation for said skills. For example, they state, ``Unless they write a senior thesis, undergraduates are also not often called upon to search the literature; read, analyze, evaluate, interpret, and cite technical articles; and make specific use of the scientific and engineering information therein, despite the fact that graduates are likely to be called upon to do so whether they pursue graduate study or enter the workforce'' \citep{heron2019joint}. Reading and writing skills may be neglected due to a number of reasons related to limitations of the instructor: lack of priority in an already content-packed curriculum \citep{williams2020first}, lack of awareness of current education research and methods \citep{van2018barriers, shkedi1998teachers, young2022using}, lack of preparation for teaching these skills \citep{adler2007reading}, and a lack of time and resources to implement new lessons \citep{henderson2007barriers,dancy2010pedagogical}. Despite these challenges, there already exist natural opportunities to explicitly introduce reading and writing skills into physics and astronomy curricula, e.g. via lab courses and CUREs. 

Some prior work on reading comprehension in the science classroom has been completed in physics, including the use of metacognition (i.e. reflection about thought processes) and students' question formulation for improving reading skills \citep{koch1991improvement,koch2001training}. However, this research is generally limited, hyper-focused on a specific reading strategy, and not grounded in authentic scientific research literature. The available literature on reading comprehension in astronomy-specific courses is even more limited (e.g. one study on reading for introductory non-major courses \citep{garland2007using}). In addition, the unique culture and demographics of each STEM sub-field (i.e. physics being substantially less gender balanced, \citep{baram2011quantifying}) prevents us from blindly extrapolating from the research of other disciplines (such as biology or chemistry where research comprehension has been more commonly studied with findings that support the importance of these skills, e.g. \citep{susiati2018correlation}), due to the substantial differences in the target population. 

Research literature can also be made more accessible via simplified summaries written by experts in the field \citep{kohler2018aas,young2022using}. Accordingly, offering easy-to-implement, readily available \rev{assignments} featuring both reading comprehension skills and accessible summaries of recent literature may be beneficial to student skills and confidence with research literature, as well as the uptake of educators actually implementing these \rev{assignments}. 

The concept of using accessible summaries (specifically ``science bites'') articles for reading comprehension exercises in the classroom was originally presented in \revv{Sanders, Kohler, and Newton 2012} \citep{sanders2012preparing} while practical infrastructure, in the form of sample \rev{assignments}, is provided in \revv{Sanders et al. 2017} \citep{sanders2017incorporating}. Building from these works, this investigation aims to gauge the efficacy of interventions focused on reading comprehension and communication skills involving simplified current literature summaries in achieving the following goal: improving student confidence with and comprehension of research literature. We propose that accessible summaries of research literature can be a powerful step towards improving students' skills and comfort with primary research papers.

We use a \rev{reading comprehension assignment} from the Astrobites website\footnote{\href{https://astrobites.org}{astrobites.org}} as an example of such interventions, making this an exploratory case study from which further research can build to more generalizable conclusions; in particular, we pursue a convergent mixed methods case study, in which different types of data (qualitative, quantitative) are analyzed in conjunction for a specific sample of \revv{astronomy undergraduate} students. This reading assignment harnesses existing accessible summaries of research literature hosted on the Astrobites website supported by the American Astronomical Society (further information on Astrobites is available in Appendix \ref{appendix:astrobites}). In this work, we distributed the aforementioned \rev{assignment} and oversaw its incorporation into undergraduate astronomy courses during the 2022-2023 academic year. 

Although existing resources from Astrobites can be adapted for a variety of educational levels, this study focuses on the applications of \rev{this assignment} to undergraduates who are enrolled in an astronomy or related majors\footnotetext{Related majors include physics, astrophysics, biology, math, geology, engineering, chemistry, computer science, planetary science, and undeclared majors.}. We seek to evaluate the effects of the \rev{assignment} on students' perceived ability to engage with astronomy literature (particularly with respect to parsing jargon, understanding the main takeaways of a paper, their conceptual understanding/intuition, and their ability to communicate about science) and their perceptions of their broader ability and belonging in the field of astronomy.

\rev{In Section \ref{methods}, we detail the reading assignment intervention, the design of the assessments used in this study, and the study population. In Section \ref{quant}, we present the analysis and results of the quantitative (Likert scale) survey data, and in Section \ref{qual}, we present the analysis and results of the qualitative survey responses from students and interviews with educators. The qualitative and quantitative data are brought together in a consistency analysis in Section \ref{consistency}.} Finally, in Sections \ref{discussion} and \ref{conclusions}, we explore the impacts of this assignment as shown in our results and suggest future uses and investigations inspired by our findings.

\section{Methods}\label{methods}

\subsection{Intervention}\label{lessonplans}

\rev{The Astrobites reading assignment template (referred to as ``Lesson Plan Type 1: Periodic Astrobites Reading Assignment'' in the Astrobites resources) was originally developed and presented in \citet{sanders2012preparing} and \citet{sanders2017incorporating}. In this assignment, students are asked to read an assigned Astrobites article and respond to guided questions that test reading comprehension and conceptual understanding. Students are graded on their responses to the questions, which can be gathered electronically through an online form. The questions are discussed in class to promote greater understanding. }

\rev{The Astrobites assignment template provides instructions and guidelines for adaptations to instructors, but leaves the specific Astrobite and reading questions to the instructor implementing the lesson. This assignment template has already been in use for over five years by educators across the U.S., and educators in this study voluntarily opted in to adopting the study intervention based on the merits of the materials available. Three other assignments that use Astrobites summaries are available (a research project, writing project, and presentation -- see Appendix \ref{assignments-info} for details), but this study is limited to the reading assignment.} 

\subsection{Assessment Design}\label{assessment}

All questions, both qualitative and quantitative, were designed specifically for this study. Assessments were developed in collaboration with UCLA's Center for the Integration of Research, Teaching, and Learning (CIRTL), and drew on established assessments from education research literature as inspiration to create the surveys \citep{bartlett2018astronomy,elby2001helping,estrada2011toward,espinosa2019reducing}. The survey questions were designed to fit into six conceptual categories \rev{of interest to the research team, based on the stated learning objectives of the assignment templates}: perception of ability to parse jargon (abbreviated as ``jargon''), \rev{perception of }ability to extract main takeaways (``main takeaways''), \rev{perception of }ability to understand astronomical concepts (``conceptual''), \rev{perception of }ability to communicate science (``communication''), \rev{perception of }general ability in the field (``ability''), and sense of belonging (``belonging'') as summarized in Table \ref{tab:conceptcats}. Each question in a category is designed to probe a similar underlying concept.  The assessments were tested with astronomy graduate students, who likely have an experience level intermediate to the undergraduates experiencing the lessons and the instructors administering them, to ensure the question text was interpreted reliably and as expected \citep{litwin1995measure,taherdoost2016validity}; only minor wording revisions were needed as a result of this testing, and revisions were decided based on \rev{think-aloud interviews} with the testing participants where we asked testing participants to verbalize their thoughts in response to question text \citep{pepper2018think,wolcott2021using}.

Student surveys were administered by participating instructors before and after implementing the \rev{assignment} in the form of a Google Forms survey. The survey contained both Likert-scale quantitative questions and open-ended written response questions. The qualitative and quantitative data collection for this study took place in parallel. Students responded to both types of questions in a single survey. The investigation is a convergent mixed-methods study, wherein qualitative and quantitative data are analyzed separately and then synthesized to provide a broader view of the problem. As stated in \revv{Ponce and Pag\'an Maldonado 2015} \citep{ponce2015mixed}, ``The quantitative approach measures the objective aspects of the problem and the qualitative phase enters the subjective aspects of the problem or the experiences of the participants.'' A consistency analysis was also completed between the results of the qualitative and quantitative data collected for three quantitative items to establish validity.

\revv{Since} the survey did not directly assess student ability, only their \textit{perceptions} of their ability, we cannot make strong claims about changes in their actual abilities or skills, only their confidence. Post-intervention surveys also included direct questions about \rev{their experience and the quality of the assignment} itself. Student responses were not anonymous, but identifying information was only used to match \rev{pre-survey} and post-survey responses and categorize students to their instructor. Additionally, instructors participated in post-lesson interviews to gather qualitative information about the efficacy of the \revv{interventions} and avenues for future improvements. All study protocols and assessments described within were approved by both UCLA and University of Virginia's Institutional Review Boards (UCLA IRB\#22-001473 and UVA IRB\#5514). Full text of the assessment questions is available in Appendix \ref{survey-app}.

\rev{We measured the internal consistency of our assessment using Cronbach's alpha, which tells us the degree to which the questions in each of the six conceptual categories are closely related. This statistic is commonly used in physics education research \citep{formanek2019, PhysRevPhysEducRes.20.020107, PhysRevPhysEducRes.20.020141}. An acceptable value for Cronbach's alpha is $>0.7$, with a value $>0.9$ considered excellent \citep{Cronbach1952}. We calculated Cronbach's alpha using the \revv{same dataset from 52 students as used in the remainder of the analysis, described further in the following section.} Table \ref{tab:conceptcats} gives the Cronbach's alpha values for each conceptual category, as calculated for the \rev{pre-survey} and post-survey responses separately.} \rev{As the Cronbach's alpha values all fell within acceptable limits, we felt comfortable proceeding with this model of the conceptual categories for our analysis.} 

\begin{table*}[!ht]
\caption{\rev{The numbers of questions included in each of the conceptual categories. The text of each question can be found in Section \ref{survey-app}. Each category is presented with its Cronbach's alpha values, which provide a measure of internal consistency for the six conceptual categories in the survey.}
\label{tab:conceptcats}
}
    \centering
    \begin{tabular}{|l|c|c|c|}
    \hline
        Category & Question numbers in category & Pre-Survey ($N=52$) & Post-Survey ($N=52$) \\ \hline \hline
        Jargon & 1-5 & 0.776 & 0.743 \\ \hline
        Main Takeaways & 6-11 &  0.915 & 0.878\\ \hline
        Conceptual & 12-18 &  0.925 & 0.885\\ \hline
        Communication & 19-24 &  0.937 & 0.928\\ \hline
        Ability & 25-29 & 0.888 & 0.857 \\ \hline
        Belonging & 30-35 & 0.906 &0.897 \\ \hline
    \end{tabular}
\end{table*}

\subsection{Recruitment \& Study Population}\label{studypop}

The data used in this study is a subset of a larger dataset including all four Astrobites-related assignments. \rev{Although the whole dataset includes 203 unique students, our sample focuses on the reading assignment \textit{only} and contains 52 students from seven courses for astronomy-related majors taught by six instructors across six institutions from Fall 2022 to Winter/Spring 2023 who completed both the pre-survey and the post-survey.} Student participants were undergraduates with an intention to major in astronomy or a related subject, and were either taking the course featuring the \rev{assignment} from Astrobites for major or minor credit, or intending to pursue an astronomy-related career. \rev{Five of the host institutions (hosting six of the courses) were R1  universities \revv{(doctoral-granting institutions with a high level of research activity)}, while one was a liberal arts college (hosting one of the courses). We also note that our sample size is small in part due to the fact that undergraduate astronomy majors are a small population -- 2023 AIP reports list $\sim20,000$ enrolled juniors and seniors in physics, but only $\sim2,800$ enrolled in astronomy \citep{AIP_2024a,AIP_2024b}.} 

\rev{Participating instructors were recruited via a public call for participants, posted online on the Astrobites website and emailed to faculty and institutional email lists. No compensation was offered. Instructors were allowed to choose from any of the available Astrobites assignment templates, although the data analyzed in this work only includes those who chose the reading assignment.} All instructors were required to participate in a pre-intervention meeting to ensure knowledge of study protocols (including how to distribute student survey assessments). Also required was a post-intervention meeting to gather information on the instructors' experiences incorporating the \revv{interventions} in their classes. Each participating instructor corresponded with the research team to implement the assignment in their classroom; research team members only advised the instructor and answered questions, but did not have any direct contact with the students themselves. \rev{Students were notified of the surveys to fill out by their instructor, using the instructor's preferred course communication platform.} Instructors notified the study team of any deviations from the assignment template as provided. Some instructors were allowed to repeat an \rev{assignment} multiple times and relevant data was collected, although the analysis of dosing effects in those data \citep{zhai2010dosage} from exposure to multiple lessons is not included in this work \revv{and we do not include students' repeated exposures in this dataset}.

\section{Quantitative Analysis \& Results}\label{quant}

The data set for quantitative analysis consists of 52 complete \rev{pre-survey} and post-survey responses to the 35 Likert-scale questions listed in Appendix \ref{survey-app}. We summarize the change in the responses to each question with the 95\% confidence interval (`CI95\%') for the mean of the \rev{pre-survey/post-survey} differences in student responses. Using the statistics package \texttt{Pingouin} \cite{Vallat2018}, we perform a paired t-test for each question by pairing \rev{pre-survey} and post-survey responses for individual students. For each question, the p-value  was determined using the canonical significance threshold of 0.05 \citep{bakan1966test}.

The power is the probability of correctly rejecting a false null hypothesis `H0'. In this study, higher power (values closer to 1) indicates a lower risk of incorrectly finding no change between the \rev{pre-lesson and post-lesson} surveys. The Bayes factor in favor of the alternative hypothesis (`BF10') (e.g., \citep{bartos2023general}) is computed and \revv{expresses the ratio of the likelihood of the data} under the alternative hypothesis, `H1': that student perceptions changed\revv{, to the null hypothesis, `H0': that student perceptions did not change.} A BF10 value of 1 demonstrates equal evidence for the null and alternative hypotheses. Values above 1 up to 3 constitute anecdotal evidence, from 3 to 10 show moderate evidence, from 10 to 100 show strong to very strong evidence, and values over 100 are considered extreme\revv{, as defined in the \citet{lee2014bayesian} scale.} \revv{BF10 values less than 1 represent (varying) levels of evidence for the null hypothesis, with smaller values representing stronger evidence for H0 -- that is, evidence that there was \textit{no} effect.}

\begin{figure*}
    \centering
    \includegraphics[width=0.65\linewidth]{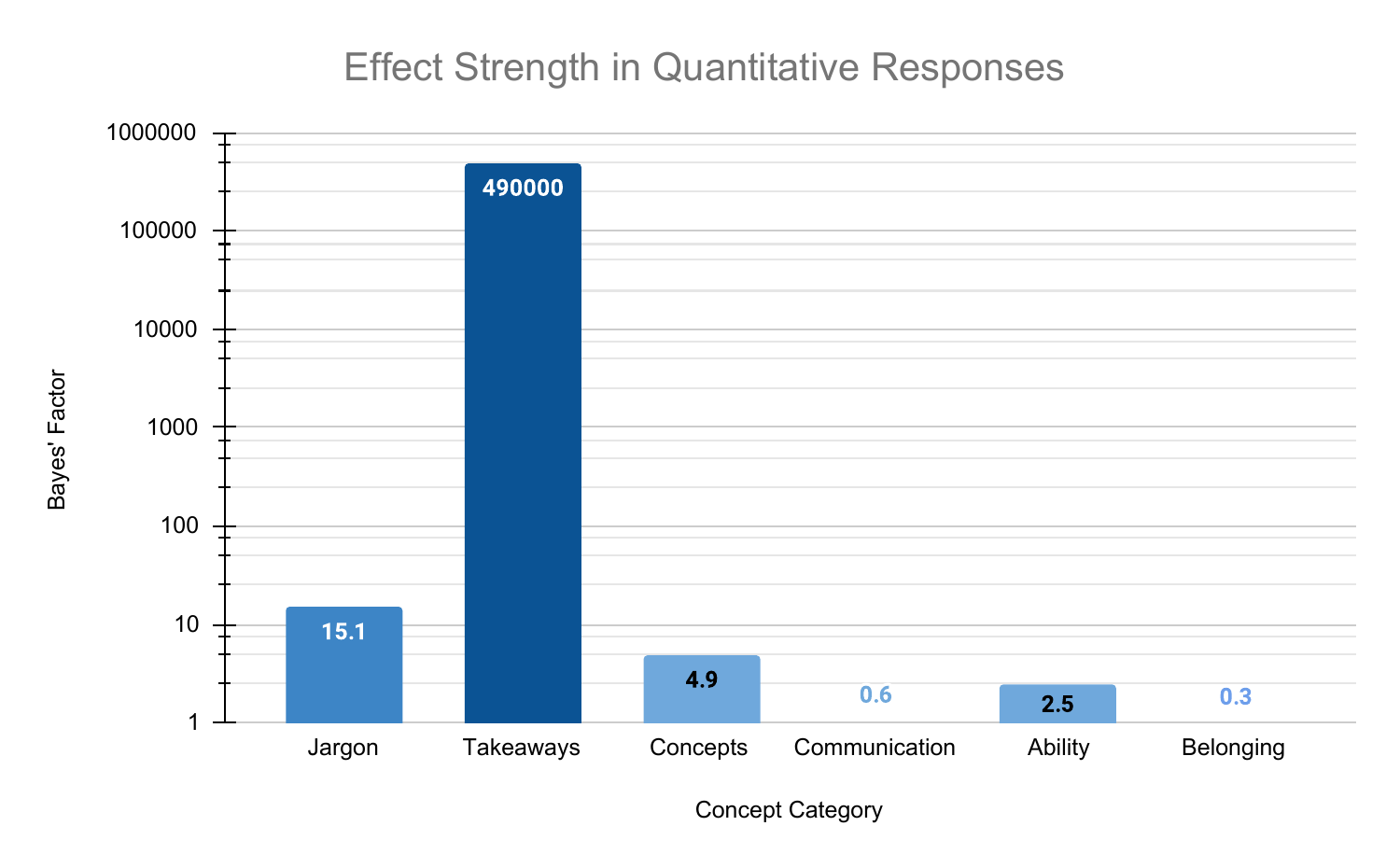}
    \caption{Visualization of effect strength across the six categories of concepts probed in the survey assessment, as presented in Table \ref{tab:groupstats}. Effects are colored by their ``strength'' of the Bayes' Factor interpretation, from anecdotal (lightest blue) to extreme (darkest blue).}
    \label{fig:quant-categories}
\end{figure*}

We also analyzed changes within the six designed categories; the same statistical analyses performed with responses to individual questions were applied to each of the conceptual categories. ``Reversed'' Likert scale questions were adjusted\footnote{For quantitative questions 2 and 5, low Likert scores (disagreement) represent \textit{greater} confidence. This is the opposite of the majority of the questions, for which higher Likert scores indicate greater confidence.}
by reversing the Likert scale for these responses prior to performing any group calculations. Summaries from the quantitative analysis are shown in Table \ref{tab:groupstats} and Figure \ref{fig:quant-categories} for the conceptual categories. 

\begin{table*}[]
\caption{Student perceptions for the groups of questions as described in Section \ref{methods}. Sample size = 52. CI95\% is the 95\% confidence interval of the mean change, BF10 is the Bayes factor of the alternative hypothesis, and power is the probability of correctly rejecting the null hypothesis, as described in Section \ref{quant}. \label{tab:groupstats}
}
\begin{tabular}{|l|l|l|c|c|c|l|}
\hline
Group           & p-value       & CI95\%     & BF10 & BF10                       & power       \\ 
                &                  &                &       & interpretation            &               \\ \hline \hline
Jargon          & \revv{$<0.001$}& {[}0.10 0.39{]}  & 15.1  & strong                     & 0.70203 \\ \hline
Main Takeaways  & \revv{$<0.001$} & {[}0.25 0.51{]}  &  $4.91 \times 10^{5}$  & extreme    & 0.99923 \\ \hline
Conceptual      & \revv{$<0.01$} & {[}0.05 0.29{]}   & 4.9 & moderate                 & 0.69154 \\ \hline
Communication   & \revv{0.035} & {[}0.01 0.24{]}   & 0.6   & anecdotal for H0                    & 0.36436 \\ \hline
Ability         & \revv{$<0.01$} & {[}0.06 0.36{]}    & 2.5 & anecdotal                & 0.67631 \\ \hline
Belonging       & \revv{0.075} & {[}0.01 0.25{]}  & 0.3 & anecdotal for H0     & 0.29360 \\ \hline
\end{tabular}
\end{table*}

Almost all questions showed positive 
mean of differences, indicating an increase in Likert score, corresponding to an average increase in perceived abilities between the \rev{pre-survey} and post-survey. 
The exceptions are questions 14 and 17 (``I am able to relate different astronomy concepts to each other.'' and ``I am able to draw analogies between concepts discussed in astronomy literature and concepts prevalent in everyday life.''), which showed a mean difference of 0, and questions 26 and 34 (``I am capable of succeeding in my astronomy courses.'' and ``I have a strong sense of belonging to the community of astronomers.''), which each showed an insignificant ($p\sim1\gg0.05$) decrease (0.01 points each) in Likert score.

The analysis for the conceptual categories of questions is shown in Table \ref{tab:groupstats}. Every category except `Belonging' showed significant change \revv{(interpreted from the calculated p-values)} from \rev{pre-lesson} to post-lesson; BF10 provided evidence in favor of change for all groups except `Communication' 
and `Belonging'. Note that these two categories have low power and thus true effects may not be detected. Notably, there is extremely strong evidence for change in the group's perceived ability with main takeaways. There is strong evidence in change in perceived ability with jargon and moderate evidence for change in perceived ability with conceptual understanding/intuition. Evidence for changes in perception of ability within astronomy was only anecdotal. There was a preference for no change in perception of science communication ability and feelings of belonging within astronomy.

Students improved to some degree across all categories, with the highest degree of change in student perception of ability to extract main takeaways; it is worth reiterating that these are student perceptions of their ability with primary research articles, as affected by their engagement with accessible research summaries.

\section{Qualitative Analysis \& Results}\label{qual}

We explored student responses to open-ended survey questions regarding perception of ability with research \& literature (Questions 36, 38, 41, and 44 in Appendix \ref{survey-app}) through \rev{thematic analysis \citep{clarke2015thematic}. Specifically, we use the coding reliability approach for thematic analysis, wherein we aim for consistency between coders. These four questions were chosen for their relevance to the learning outcomes of the reading comprehension assignment template.}  

A set of thematic categories (referred to herein as codes, summarized in Table \ref{table:codes}) were determined inductively for each qualitative question, based on the content of student responses. All responses were read before creating codes, and then responses were re-read several times to refine the initial set of themes/codes. \rev{After collective creation of these codes by the study team, seven raters (authors BLL, NKC, RRL, SLW, SG, MD and collaborator AM) independently assigned codes to each response. It is worth noting that three of the raters (NKC, SLW, AM) were not involved in creating the codebook; this independent group serves as a simple verification for the codes used. We only included and coded responses from the same sample as the quantitative analysis ($N=52$).}

\begin{table*}
    \caption{Open-ended student survey questions probing perceptions of ability, belonging, and learning from the \revv{interventions}, along with the codes used to categorize them by theme. Examples are given in parentheses. }\label{table:codes}
\begin{tabular}{|p{1.5in}|p{5in}|}
\hline
Survey Question & Codes \\ \hline \hline
Q36: How do you feel about reading scientific research papers? & \textbf{Intimidation} (intimidated, daunting, overwhelmed, confused, not comfortable); \textbf{Require effort} (time-consuming, tedious, read multiple times); \textbf{Worthwhile} (rewarding, useful, necessary); \textbf{Unprepared} (lacking knowledge, not taken enough classes, ``not ready yet"); \textbf{Improved} (e.g., ``better now"); \textbf{Confidence} (comfortable, good, ``like I can do it"); \textbf{Enjoyment}(interesting, fun, enjoy)   \\ \hline
Q38: Do you think you are capable of being an astronomer (or related scientist)? Why or why not? & \textbf{Yes; No; Maybe} (depends, yes and no); \textbf{Confident} (good at it, previous or current experience, have what it takes, success in classes); \textbf{Insecure/uncertain} (no, maybe, unsure, grades, imposter syndrome, doubt); \textbf{Passion} (life long interest, plans for graduate school)   \\ \hline
Q41: What is your main takeaway from this activity? & \textbf{Astronomy knowledge} (topics, interpreting plots, data); \textbf{Understanding of field} (careers, breadth of topics, better, how papers are written, how science works and results are communicated, programming, statistics, visuals); \textbf{Confidence} (boost, more capable, more confident, less daunting, more daunting, decreased); \textbf{Accessibility} (more accessible, digestible, resources. importance of good teaching and/or accessible literature); \textbf{Decreased confidence} (more daunting, decreased)  \\ \hline
Q44: How have your feelings about astronomy research changed as a result of this activity, if at all? & \textbf{Improved} (more capable, less daunting, more confident); \textbf{Enjoyment} (appreciation for sci. comm., more inspired to do astro., more excited to read papers); \textbf{Meta-knowledge of field} (awareness of resources, knowledge of career options, accessibility of field); \textbf{No change}  \\ \hline
\end{tabular} 
\end{table*}

\rev{Due to the inherently subjective nature of coding, we assessed inter-rater agreement using the full set of codes applied by the seven independent coders. While there are a range of statistics used to determine inter-rater agreement and reliability, many of the more commonly used ones (i.e. kappa indices, intraclass correlation coefficient) do not apply to our data wherein we have more than two raters and the ability to apply multiple codes to each response \citep{GISEV2013330}. We instead elected to use the proportion agreement, which we calculate by taking the number of codes for each response that all raters agreed on (either all raters applying the code, or all raters not applying the code) and dividing by the total number of codes it was possible to apply. Thus if there were six codes for a given question, the proportion agreement would be equal to $(N_7+N_0)/6$ where $N_7$ is the number of codes all seven raters selected for the response and $N_0$ is the number of codes none of the raters selected. This statistic does not take into account the possibility of chance agreement, making it slightly less robust, but given the large sample size and number of coders, as well as the lack of mutual-exclusivity among the codes, we determined this was the best way to proceed. There is a precedent of using proportion agreements in these types of cases \citep{Grayson2001, Campbell_2013, hewitt2023development}. The resulting percentage agreement for each question is summarized in Table \ref{tab:prop_agreement}. In general, while there is no consistently-used cutoff for an acceptable level of agreement, cutoffs generally fall between 0.7 and 0.8, and here we consider a proportion agreement $>0.7$ to be acceptable \citep{Campbell_2013, Stemler2004ACO, kurasaki, HOLDFORD2008173}.} \revv{We calculated these proportion agreements for each question, not per code, due to the nature of our codebook--that is, each question had its own unique codes, and multiple codes were able to be applied to a single response.}
\begin{table}[ht]
    \centering
    \begin{tabular}{|c|c|}
    \hline
        Question Number  &  Proportion Agreement \\ \hline \hline
        36 & 0.778 \\ \hline
        38 & 0.783 \\ \hline
        41 & 0.717 \\ \hline
        44 & 0.815 \\ 
        \hline
    \end{tabular}
    \caption{\rev{Proportion agreement metric for inter-rater reliability in our qualitative data. All questions fell within the $>0.7$ limit. }}
    \label{tab:prop_agreement}
\end{table}

\rev{We found that all of the questions used in this analysis fall within the limit of $>0.7$, although Q41 is near the lower end. Upon a brief examination of causes of inconsistency, we found that for responses that did not have complete agreement, agreement levels were still generally high (i.e. 6/7 coders applying a given code). There were no instances where the same response yielded contradicting codes (i.e. both `Yes' and `No' for Q38, or both `Improved' and `No change' for Q44). More often, differences in code applications arose in unpacking the reasoning of participant's answers. The codes `Understanding of field' in Q41 and `Confidence' in Q36 saw the largest amount of disagreement in applications between coders. These were codes that were usually applied in tandem with other codes, as responses coded `Understanding of Field' were also often coded with `Accessibility', and those coded with `Confidence' also had many coders selecting `Passion'. This suggests that individual coders may be drawing some connections between codes not explicitly designated in the codebook. Despite these few instances of lower agreement, the generally strong proportion agreement values allow us to proceed with analyzing these code applications.  }

It is worth noting that some students left questions blank, or provided responses that were illegible or unable to be matched to a code. Qualitative response codes were tallied, and the \rev{number} of responses in each main code category (e.g., not including sub-codes) was compared between \rev{pre-lesson} and post-lesson surveys. For our analysis, codes applied by a majority of raters \rev{($\geq4$)} were included as the final set of identified codes for each response, essentially determining the interpretation of the response by \revv{majority}. Codes applied by \rev{$\leq3$} raters were not included. 

\subsection{Student Responses}\label{studentqual}

\rev{The raw counts and total differences in number of responses assigned a given code are shared in Tables \ref{paperstab}, \ref{capabletab}, and \ref{postqs}. In this section, we summarize the notable findings for each qualitative question.}

\begin{table*}[]
\begin{tabular}{|lllllll|}
\hline
\multicolumn{7}{|l|}{Q36: How do you feel about reading scientific research papers? }                                                                            \\ \hline
\multicolumn{1}{|l|}{}     & \multicolumn{1}{l|}{Intimidation} & \multicolumn{1}{l|}{Require Effort} & \multicolumn{1}{l|}{Worthwhile} & \multicolumn{1}{l|}{Unprepared} & \multicolumn{1}{l|}{Confidence} & Enjoyment \\ \hline
\multicolumn{1}{|l|}{Pre}  & \multicolumn{1}{l|}{\revv{1}1}  & \multicolumn{1}{l|}{11}  & \multicolumn{1}{l|}{7}     & \multicolumn{1}{l|}{9}        & \multicolumn{1}{l|}{10}      & 17                 \\ \hline
\multicolumn{1}{|l|}{Post} & \multicolumn{1}{l|}{14}  & \multicolumn{1}{l|}{8}  & \multicolumn{1}{l|}{8}     & \multicolumn{1}{l|}{2}        & \multicolumn{1}{l|}{19}      & 17                  \\ \hline
\multicolumn{1}{|l|}{Difference} & \multicolumn{1}{l|}{-7}   & \multicolumn{1}{l|}{-3}  & \multicolumn{1}{l|}{1}    & \multicolumn{1}{l|}{-7}        & \multicolumn{1}{l|}{9}      & 0                 \\ \hline
\end{tabular}\caption{Codes applied to responses to Question 36 probing students' feelings around reading research papers. There was an additional code for students who explicitly claimed it had improved, but it was never applied by four or more coders to the same response. Overall, more negative codes (intimidation, unprepared, require effort) saw a decrease in application in the post-survey, while codes indicating students had confidence or felt it was worthwhile increased. }\label{paperstab}
\end{table*}

\begin{table*}[]
\begin{tabular}{|lllllll|}
\hline
\multicolumn{7}{|l|}{Q38: Do you think you are capable of being an astronomer (or related scientist)? Why or why not?}                                                                            \\ \hline
\multicolumn{1}{|l|}{}     & \multicolumn{1}{l|}{Yes} & \multicolumn{1}{l|}{No} & \multicolumn{1}{l|}{Maybe} & \multicolumn{1}{l|}{Confident} & \multicolumn{1}{l|}{Passion} & Insecure/Uncertain \\ \hline
\multicolumn{1}{|l|}{Pre}  & \multicolumn{1}{l|}{42}  & \multicolumn{1}{l|}{1}  & \multicolumn{1}{l|}{8}     & \multicolumn{1}{l|}{29}        & \multicolumn{1}{l|}{18}      & 12                 \\ \hline
\multicolumn{1}{|l|}{Post} & \multicolumn{1}{l|}{45}  & \multicolumn{1}{l|}{2}  & \multicolumn{1}{l|}{4}     & \multicolumn{1}{l|}{22}        & \multicolumn{1}{l|}{17}      & 7                  \\ \hline
\multicolumn{1}{|l|}{Difference} & \multicolumn{1}{l|}{3}   & \multicolumn{1}{l|}{1}  & \multicolumn{1}{l|}{-4}    & \multicolumn{1}{l|}{-7}        & \multicolumn{1}{l|}{-1}      & -5                 \\ \hline
\end{tabular}\caption{Responses to Question 38 probing students' perceptions of their capability in astronomy. 6 students reported improvement and 3 students reported a decline in feelings of capability, but most responses entered positive/affirmative and remained so.}\label{capabletab}
\end{table*}

\begin{table*}[]
\begin{tabular}{|lllll|}
\hline
\multicolumn{5}{|l|}{Q41: What is your main takeaway from this activity?}                                                                      \\ \hline
\multicolumn{1}{|l|}{Knowledge} & \multicolumn{1}{l|}{Field Understanding} & \multicolumn{1}{l|}{Confidence}     & \multicolumn{1}{l|}{Accessibility} & Decreased confidence \\ \hline
\multicolumn{1}{|l|}{9}         & \multicolumn{1}{l|}{14}                  & \multicolumn{1}{l|}{7}              & \multicolumn{1}{l|}{13}            & 3              \\ \hline
\multicolumn{5}{|l|}{Q44: How have your feelings about astronomy research changed as a result of this activity, if at all?}                    \\ \hline
\multicolumn{1}{|l|}{Improved}  & \multicolumn{1}{l|}{Enjoyment}           & \multicolumn{1}{l|}{Meta-knowledge} & \multicolumn{2}{l|}{No Change}                      \\ \hline
\multicolumn{1}{|l|}{11}        & \multicolumn{1}{l|}{7}                   & \multicolumn{1}{l|}{5}              & \multicolumn{2}{l|}{25}                             \\ \hline

\end{tabular}\caption{Student response counts for Questions 47 and 44, with students' self-reported takeaways and changes in attitudes from the intervention. For students who reported some change int heir attitudes towards astronomy research, all changes were positive.}\label{postqs}
\end{table*}

\rev{To analyze the responses to Question 36 (``How do you feel about reading research papers?'') and Question 38 (``Do you think you are capable of being an astronomer?''), we matched students' \rev{pre-survey} and post-survey to see how their responses changed over time. } \revv{It is worth noting that students were writing a new response without access to their pre-intervention survey for the post-intervention survey, so they may have shared different information than on the initial survey; this may account for some of the responses that contained a certain code in the pre-intervention survey but not in the post-intervention survey.}

\begin{figure*}
    \centering
    \includegraphics[width=\linewidth]{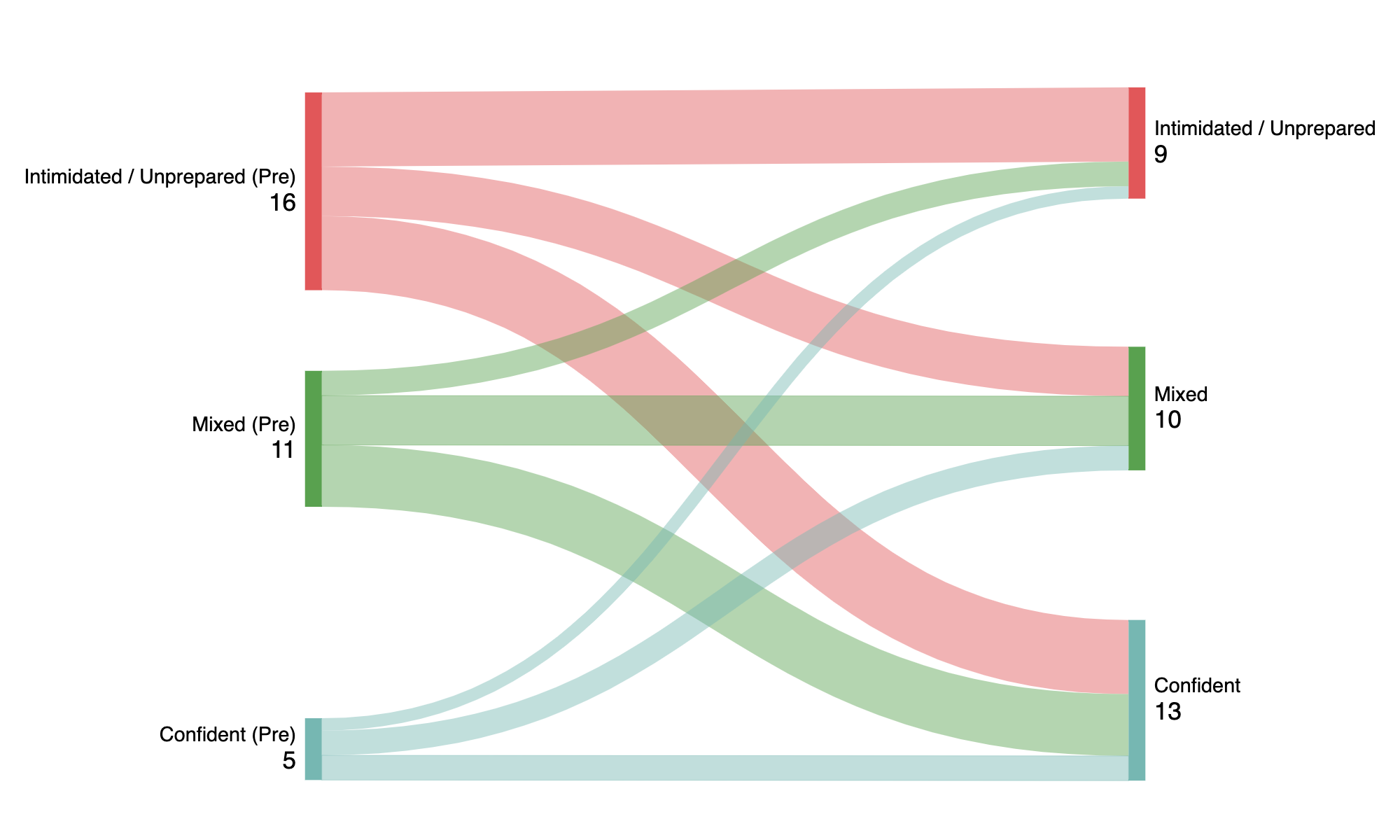}
    \caption{Changes in individual student responses to Question 36, ``How do you feel about reading research papers?'' for students who mentioned intimidation, unpreparedness, and/or confidence in either their \rev{pre-lesson} or post-lesson surveys. Responses that included the codes intimidated and/or unprepared are categorized in red, while responses that included the code confidence are categorized in blue. The ``mixed'' category includes responses that (1) did not have codes for confidence, intimidation, or unpreparedness or (2) had both confidence and improvement or unpreparedness indicated. A majority of students improved (e.g. intimidated to mixed feelings, mixed to confident, or intimidated to confident) although some students did report more negative feelings after the lesson.}
    \label{fig:sankey36}
\end{figure*}

\rev{For Q36, 20 of the pre-survey and \revv{16} of the post-survey responses indicated feeling \revv{intimidated by or unprepared for} scientific papers. There were eight students whose comments suggested feeling intimidated in the pre-survey, but did not in post, although four students were not coded for being intimidated for pre-, but were for post-. It is worth noting that for three out of those four, their post-survey responses also indicated something positive about the experience, as one said ``They’re a little daunting, but I usually enjoy it." There were also eight students who felt unprepared in the pre-survey, but did not mention that concern in the post, and one explicitly stated ``Reading astronomy papers has gotten easier with time." However, four of those eight did still indicate that they were intimidated in their post-survey responses, suggesting they weren't yet fully comfortable with reading papers even if there was a slight improvement. Many students also stated that they enjoyed reading papers, with thirteen students maintaining that position in both pre- and post- responses, five saying it only in post, and five saying it only in pre. As touched on above, several of these responses came with qualifiers such as ``daunting", ``I don't always understand", and       ``sometimes the wording...is hard to follow". Finally, 10 of the 52 respondents expressed a level of confidence in the post-survey that they did not in the pre-survey, and overall 19 respondents felt at least somewhat confident about reading scientific papers at the end of the intervention. A summary of changes in individual students' reported feelings towards reading research papers is presented in Figure \ref{fig:sankey36}.}

\begin{figure}
    \centering
    \includegraphics[width=\linewidth]{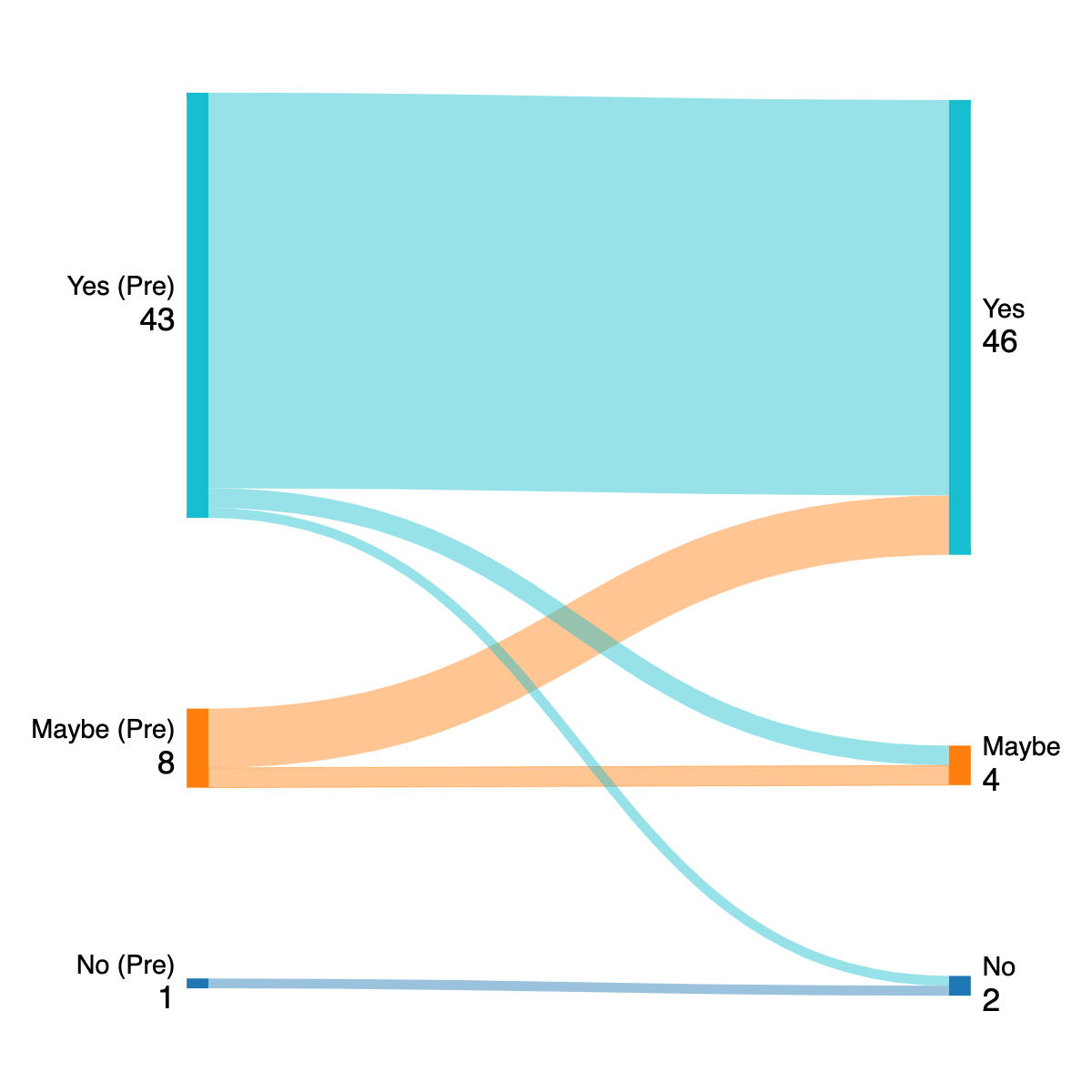}
    \caption{\rev{Sankey diagram illustrating students' changes in perception of their ability in astronomy before and after the reading intervention. A majority of students entered with a positive attitude towards their own ability, and remained so.}}
    \label{fig:sankey38}
\end{figure}

\rev{For the other paired qualitative question, ``Do you think you are capable of being an astronomer?'', a majority of students responded with yes on both the \rev{pre-lesson} and post-lesson surveys (42 pre, 45 post).  Of students whose answers changed from \rev{pre-lesson} to post-lesson, 6 reported positive change (from ``no'' or ``maybe'' to ``yes''), 2 reported a slightly negative change (from ``yes'' to ``maybe''), and 1 reported a negative change (from ``yes'' to ``no''). The student whose answer changed to no cited feelings of overwhelm in their post-lesson response: ``I don't think I have the skills necessary and feel quite overwhelmed with the way things work in astrophysics.'' Response counts for Question 38 are provided in Table \ref{capabletab}, and changes in student responses for Question 38 are illustrated in Figure \ref{fig:sankey38}.}

\rev{Two questions were available on the post-lesson survey only. 
For Question 41 ``What is your main takeaway?'', we categorized student responses into five categories: astronomical knowledge gains (9 students), improvements to students' understandings of the field/culture of astronomy (14 students), improved confidence (7 students), decreased confidence (3 students), and a greater awareness of the possibilities/need for accessibility in the field (13 students). For the respondents who shared a decreased confidence, the largest cause seemed to be difficulty in reading papers due to the amount of jargon and technical language. As one student put it ``Astronomy papers are filled with a lot of information only meant for astronomers.'' For those students who had increased confidence, some things they identified that helped them were having a general idea of the scientific background and taking their time with reading. Of the 13 respondents coded for `Accessibility', a common trend in the responses was how the key takeaways from research can be presented more simply than what is seen in papers. As one student explained, ``Scientific findings can be phrased in relatively simple ways, making the findings of research more accessible to  people who don't already know a lot about the material.'' }

\rev{For Question 44 ``How have your feelings about astronomy research changed?'', eleven students indicated that their feelings improved, 25 indicated that there was no change, and no students reported more negative feelings about astronomy research after the activity. We additionally coded responses for mentions of enjoyment and meta-knowledge of the field, with seven respondents mentioning enjoyment and five gaining more meta-information. Among the meta-knowledge responses, one student mentioned they now knew how to find more resources, while the others focused around more general career advancement, as one respondent shared ``I have a better understanding about my future in the field." 
Response counts for Questions 41 and 44 are also summarized in Table \ref{postqs}.}

\subsection{Instructor Feedback}\label{educatorqual}

\rev{Due to the small number of instructors, we consider their feedback on all assignment templates implemented, including but not limited to the reading assignment.} All instructors said the \rev{assignment template they implemented} was well-received in their class, and mentioned the following benefits of the \rev{assignments}: helping to get students reading papers, exposing them to what kind of research exists in the field, increased engagement in class, increases in student confidence, awareness of Astrobites as a resource, and a change of pace from a normal routine (e.g., problem sets).

Instructors nearly unanimously stated the most useful aspect of the \rev{assignment template} was simply that it exists; structured/pre-made \rev{resources} connecting to current research that can be easily slotted into a course are an extremely beneficial resource for already over-committed educators. 

Instructors additionally provided helpful feedback and suggestions for improvement, which we will take under consideration next time we revise the \rev{assignment template} resources. 
These improvements include: incorporating ``classic'' (e.g., foundational or historical) papers for their pedagogical value and fundamental roles in the field and adding a structured option to do the reading assignment as a recurring activity.
Additional illustrative quotes from educator qualitative responses are available in Appendix \ref{app:quotes}.

\section{Consistency Analysis}\label{consistency}

When performing a mixed methods study such as this, it is good practice to compare results obtained through multiple questions covering the same themes. One way to conduct such a comparison involves assigning numerical scores to qualitative questions 
\cite{Caracelli1993,Onwuegbuzie2010}. 

Here, we identified three qualitative questions (Questions 36, 38, and 39) for which there are Likert scale questions that address similar themes. We determined which qualitative response codes denoted positive, neutral, and negative feelings toward the three questions, and assigned each student a qualitative ``score'' for that question based on the prevalence of each type of code in their response.
The codes considered positive, neutral, and negative for each of the three qualitative questions used in this comparison are listed in Table \ref{tab:quan-qual}.

\begin{table*}[]
    \centering
    \caption{Details of qualitative-quantitative consistency analysis. Question 36: ``How do you feel about reading scientific research papers?'' Question 38: ``Do you think you are capable of being an astronomer (or related scientist)? Why or why not?'' Question 39: ``Do you feel like you belong in astronomy? Why or why not?'' In figures showing results of this consistency analysis, these are referred to as ``Papers,'' ``Capable,'' and ``Belonging.'' \label{tab:quan-qual}}
   \begin{tabular}{|l|c|c|}
        \hline
        Open-Ended Question & Codes considered & Related Likert questions \\ 
        \hline \hline
        36 (Papers) & Intimidation$(-)$, Unprepared$(-)$, Effort$(-)$, & 1, 2, 3, 4, 5, 7, 8, 9, 11, 28 \\
          & Confidence$(+)$, Worthwhile$(+)$, Enjoyment$(+)$                          & \\ \hline
        38 (Capable) & Yes$(+)$, No$(-)$, Maybe$(0)$ & 14, 18, 25, 26, 27, 28, 29, 30, 31, 32 \\ \hline
        39 (Belonging) & Yes$(+)$, No$(-)$, Maybe$(0)$ & 32, 33, 34, 35 \\ \hline
        \end{tabular}
\end{table*}

For each student response to each of these three questions, we took the average of related Likert scale questions to find their ``Likert score'' (quantitative). To assign a student's ``written score'' (qualitative) on a topic, we assigned $+1$ to every positive code applied to their response to that question, $-1$ to every negative code, and $0$ to every neutral code, \rev{then applied the average of these scores for that response. We only considered codes applied by a majority of raters, and we considered the number of raters who applied a given code. For example, if four raters applied a positive code to a response but six applied a negative code, the response received a score of $(+4 - 6)/10 = -0.2$. (Note that while we only had seven raters, each rater could apply multiple codes to the same response, and many responses included both positive and negative themes; as a result, the denominator in this calculation is often $>7$.)}
\rev{Responses to question 36 (``How do you feel about reading scientific research papers?'') often contained a mix of positive and negative codes, leading to a broader range of written scores. Questions 41 (capable) and 42 (belonging) are more straightforward (i.e., only coded with "yes," "no," and "maybe" for this analysis), and as such only have written scores of $-1$, $0$, or $1$.}



Written response scores tend to track Likert scores for each of the questions
, as computed with the Spearman correlation coefficient \rev{(where a score of 0 indicates no correlation, and $\pm1$ a monotonic correlation).} However, the correlation between the two scores is smaller in the post-survey than in the pre-survey for all three questions. \rev{(In the pre-survey, the coefficients are $0.63$, $0.55$, and $0.49$ for questions 39, 41, and 42, respectively; in the post-survey, these are $0.28$, $0.39$, and $0.40$.)} \rev{We also found that many students' quantitative scores changed more substantially than their qualitative ones; typically, their Likert scale responses were more positive while their written response scores did not change or showed only minor change.}
Additionally, students occasionally responded to Likert score questions positively (i.e., $>4$) while using neutral or negative language in the corresponding written response question. This inconsistency reflects the greater nuance available to students in written response.

\section{Discussion}\label{discussion}

Both students and educators expressed positive feelings towards the \rev{accessible summary reading assignment} in their direct feedback, and our data show positive change in all major categories probed by the quantitative survey questions (see Table \ref{tab:conceptcats}), as summarized in Section \ref{quant}. The strongest effect sizes were measured for ability with jargon and main takeaways. 

\begin{figure}
    \centering
    \includegraphics[width=1.0\linewidth]{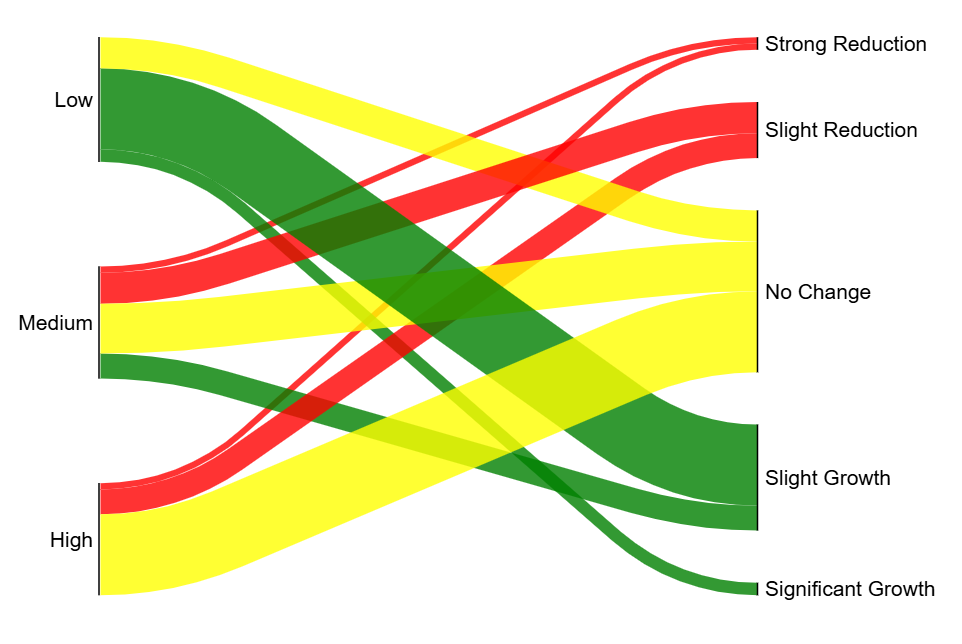}
    \caption{\rev{Sankey diagram illustrating students' changes in perception of ability to determine the main takeaways of a paper before and after the reading intervention. Floor and ceiling effects are readily apparent.}}
    \label{fig:second-order-changes-sankey}
\end{figure}

The amount of students whose scores began and remained high (as shown in Figure \ref{fig:second-order-changes-sankey} \rev{for the quantitative data, and Figure \ref{fig:sankey38} for one of the qualitative questions)} highlights a limitation in our analysis: students whose initial responses lay to one extreme or another (the ``floor and ceiling effects'' \citep{wang2008investigating}), i.e., students whose initial averages in a category were close to 1 or 7. A student with an initial average of 2 could not exhibit a decrease of more than $-1$ after experiencing the \revv{interventions}, while a student with an initial average of 6 could not exhibit an increase of more than $1$. 
\rev{Additionally, decreases or lack of change in confidence -- as in Q38 -- may also be attributed to the fact that students'} ``measuring stick'' for themselves has changed since the beginning of the course, as they have gained knowledge and practice \citep{hiebert2012assessing,Kanevsky}. 
\revv{It is also important to note that the timing of post-survey administration varied between instructors in the study (some administered it immediately after completion of the activity, some provided a window of time such as a week after the activity to complete the survey on their own time), and this may have had an effect on student responses, providing them more time in which to gain said knowledge and practice. Furthermore, }we only completed the treatment once, and it may take repeated exposures (i.e. more practice with the \rev{assignments}) for benefits to be realized. 

Qualitative responses from students showed a more mixed and nuanced view of their development from before to after experiencing the intervention. In response to questions about feelings while reading research, students less often mentioned feelings of intimidation and unpreparedness after the lesson, while mentioning more often that papers feel important/worthwhile and they are confident in their abilities. 

For a majority of students, the main takeaway from the lesson \rev{was either a greater understanding of the field of astronomy -- e.g. how research works, what scientists do, how research publications come to be -- or an increased awareness of existing resources and the need for additional resources to support accessibility in the field.} Nearly all students' takeaways were positive and in line with the learning objectives from the \rev{assignment template}, \rev{also including increased content knowledge and increased confidence with course material. Although $\sim50$\% of students claim their feelings about astronomy have not changed as a result of the activity, those who do claim to have changed are largely positive.}

Students were generally consistent in their qualitative and quantitative responses for the pre-survey, but less so in the post-survey. 
We argue that the greater nuance available in written responses meant that students were more likely to express a mix of positive and negative feelings -- with our qualitative scoring model for consistency analysis, this could translate to more neutral or negative scores, even if their overall feelings (as reflected through a collection of Likert scale questions) showed positive change. Additionally, students and instructors were largely consistent in their feedback; instructors reported feeling that their students had gained understanding and perspective from the \rev{activities}, and many students reported either improved or reinforced\rev{confidence in their abilities.}

\section{Conclusions \& Recommendations}\label{conclusions}

Overall, our results show that the \rev{accessible summary reading assignment intervention was} largely successful at its main goal -- increasing student comfort and \revv{perceived} ability with research literature -- and yielded other beneficial effects, such as increasing student knowledge about the field of astronomy and the process of research. \rev{This intervention} significantly improved student ability and confidence with parsing jargon and understanding the main takeaways of a \revv{research summary article, which may translate to improved ability with primary research literature}. Our results, therefore, show evidence for the \rev{promise of} positive benefits of reading comprehension assignments featuring accessible summaries of current literature more generally\rev{, although it is not yet known how this intervention compares to other reading comprehension exercises.} As such, we identify a recommendation for instructors to use such assignments and a need for further research work in this area to fully maximize the benefits and understand the mechanisms at play in student learning. 

\revv{It is important to note that this study focused on a relatively small sample of students, and there is likely a bias in our sample due to the self-selecting nature of instructors who chose to participate (i.e. motivated educators seeking out new pedagogical strategies). Further work will be needed to make truly generalizable conclusions about the effects of reading comprehension interventions on student ability with research literature and other outcomes. As mentioned previously, it is also yet unclear how students' gains will change with repeated exposure to these lessons, e.g., if the majority of change is in the first exposure or if they grow substantially with practice. Future work will need to test these ``dosing effects'' to determine the optimal usage of these lessons. The sample in this paper focused on \revv{the skills necessary for the task of} reading research papers, however, there is also still significant work to be done to explore how writing lessons impact students in STEM classrooms. We envision this as complimentary work to the reading comprehension interventions we studied, especially as reading is the first step to writing \citep{barger2017reading}.}  

Open-ended responses from participants also indicate that the accessible summary was a key component of the \rev{assignment's} success in increasing student confidence with literature; scaffolding steps have long been shown to improve student outcomes \citep{van2010scaffolding}, and our work indicates a similar phenomenon for the task of reading complex scientific articles. \revv{Scaffolding students' comprehension} of scientific research articles can often be difficult, as the available alternatives to cover the content (namely science journalism) often reduce the complexity to the point of being an entirely different, non-technical genre that likely cannot target the same skills of reading technical literature. Astrobites and similar efforts represent the creation of a third writing type in the field, the accessible summary, intended for beginner technical audiences such as the students in this study population. Our results indicate that this type of writing \rev{may be} beneficial for students as they enter the realm of scientific research.

\revv{In conclusion,} accessible summaries \revv{show promise} for bridging the gap between students' entry points and the difficult task of reading scientific literature\revv{, although significant further study is needed to expand upon this exploratory case study.} Additionally, it is important to note that instructors were appreciative of the existence of pre-made resources to use, as opposed to starting from scratch to incorporate reading in their courses. Future efforts in this area should similarly work to provide instructors with usable \rev{activities and templates} to encourage uptake of reading/writing-focused activities in astronomy courses. We strongly encourage further work on integrating reading comprehension exercises and research literature into the classroom, especially through accessible summaries as in \rev{this \revv{exploratory} case study}.

\begin{acknowledgments}
B.L.L. acknowledges support from the National Science Foundation Graduate Research Fellowship under Grant No. 2021-25 DGE-2034835 and the National Science Foundation Astronomy \& Astrophysics Postdoctoral Fellowship under Award No. 2401654. A.R.W. acknowledges support from the Virginia Space Grant Consortium and the National Science Foundation Graduate Research Fellowship Program under grant No. 1842490. S.G. acknowledges support from the National Science Foundation Graduate Research Fellowship under Grant No. 2233001. Any opinions, findings, and conclusions or recommendations expressed in this material are those of the author(s)and do not necessarily reflect the views of the National Science Foundation. 

Support for C.J.L. was provided by NASA through the NASA Hubble Fellowship grant No. HST-HF2-51535.001 awarded by the Space Telescope Science Institute, which is operated by the Association of Universities for Research in Astronomy, Inc., for NASA, under contract NAS5-26555. RRL is supported by the Deutsche Forschungsgemeinschaft (DFG, German Research Foundation) in the form of an Emmy Noether Research Group -- Project-ID 445674056 (SA4064/1-1, PI Sander). This work was partially funded by the American Astronomical Society Education and Professional Development (EPD) mini-grant program.

This study was conducted with a UCLA IRB Exemption Certificate IRB\#22-001473 and UVA IRB Approval Certificate IRB\#5514. B. Lewis would like to thank the UCLA CIRTL TAR program for their guidance and mentorship, especially her CIRTL co-authors on this paper, Rachel Kennison for serving as faculty sponsor for UCLA IRB, and Graham Read. Thank you to Nicholas Young for valuable feedback on this manuscript, Jessie Thwaites and Haley Wahl for participating in talk-throughs as part of the survey design process, to Luna Zagorac for proofreading, to Nathan Sanders for his unwavering support of Astrobites' education initiatives, to Huei Sears for her support of grant proposals to fund this work, to Alexandra Masegian for her assistance in coding data, and to the rest of the Astrobites collaboration for their support of this project.
\end{acknowledgments}

\bibliography{bib}

\begin{thebibliography}{65}%
\makeatletter
\providecommand \@ifxundefined [1]{%
 \@ifx{#1\undefined}
}%
\providecommand \@ifnum [1]{%
 \ifnum #1\expandafter \@firstoftwo
 \else \expandafter \@secondoftwo
 \fi
}%
\providecommand \@ifx [1]{%
 \ifx #1\expandafter \@firstoftwo
 \else \expandafter \@secondoftwo
 \fi
}%
\providecommand \natexlab [1]{#1}%
\providecommand \enquote  [1]{``#1''}%
\providecommand \bibnamefont  [1]{#1}%
\providecommand \bibfnamefont [1]{#1}%
\providecommand \citenamefont [1]{#1}%
\providecommand \href@noop [0]{\@secondoftwo}%
\providecommand \href [0]{\begingroup \@sanitize@url \@href}%
\providecommand \@href[1]{\@@startlink{#1}\@@href}%
\providecommand \@@href[1]{\endgroup#1\@@endlink}%
\providecommand \@sanitize@url [0]{\catcode `\\12\catcode `\$12\catcode `\&12\catcode `\#12\catcode `\^12\catcode `\_12\catcode `\%12\relax}%
\providecommand \@@startlink[1]{}%
\providecommand \@@endlink[0]{}%
\providecommand \url  [0]{\begingroup\@sanitize@url \@url }%
\providecommand \@url [1]{\endgroup\@href {#1}{\urlprefix }}%
\providecommand \urlprefix  [0]{URL }%
\providecommand \Eprint [0]{\href }%
\providecommand \doibase [0]{https://doi.org/}%
\providecommand \selectlanguage [0]{\@gobble}%
\providecommand \bibinfo  [0]{\@secondoftwo}%
\providecommand \bibfield  [0]{\@secondoftwo}%
\providecommand \translation [1]{[#1]}%
\providecommand \BibitemOpen [0]{}%
\providecommand \bibitemStop [0]{}%
\providecommand \bibitemNoStop [0]{.\EOS\space}%
\providecommand \EOS [0]{\spacefactor3000\relax}%
\providecommand \BibitemShut  [1]{\csname bibitem#1\endcsname}%
\let\auto@bib@innerbib\@empty
\bibitem [{\citenamefont {Seymour}\ \emph {et~al.}(2004)\citenamefont {Seymour}, \citenamefont {Hunter}, \citenamefont {Laursen},\ and\ \citenamefont {DeAntoni}}]{seymour2004establishing}%
  \BibitemOpen
  \bibfield  {author} {\bibinfo {author} {\bibfnamefont {E.}~\bibnamefont {Seymour}}, \bibinfo {author} {\bibfnamefont {A.-B.}\ \bibnamefont {Hunter}}, \bibinfo {author} {\bibfnamefont {S.~L.}\ \bibnamefont {Laursen}},\ and\ \bibinfo {author} {\bibfnamefont {T.}~\bibnamefont {DeAntoni}},\ }\bibfield  {title} {\bibinfo {title} {Establishing the benefits of research experiences for undergraduates in the sciences: First findings from a three-year study},\ }\href {https://doi.org/https://doi.org/10.1002/sce.10131} {\bibfield  {journal} {\bibinfo  {journal} {Science Education}\ }\textbf {\bibinfo {volume} {88}},\ \bibinfo {pages} {493} (\bibinfo {year} {2004})},\ \Eprint {https://arxiv.org/abs/https://onlinelibrary.wiley.com/doi/pdf/10.1002/sce.10131} {https://onlinelibrary.wiley.com/doi/pdf/10.1002/sce.10131} \BibitemShut {NoStop}%
\bibitem [{\citenamefont {Donohue}\ \emph {et~al.}(2021)\citenamefont {Donohue}, \citenamefont {VanDenburgh}, \citenamefont {Reck},\ and\ \citenamefont {Buck}}]{donohue2021integrating}%
  \BibitemOpen
  \bibfield  {author} {\bibinfo {author} {\bibfnamefont {K.}~\bibnamefont {Donohue}}, \bibinfo {author} {\bibfnamefont {K.}~\bibnamefont {VanDenburgh}}, \bibinfo {author} {\bibfnamefont {C.}~\bibnamefont {Reck}},\ and\ \bibinfo {author} {\bibfnamefont {G.}~\bibnamefont {Buck}},\ }\bibfield  {title} {\bibinfo {title} {Integrating science communication into a large stem classroom},\ }\href@noop {} {\bibfield  {journal} {\bibinfo  {journal} {Journal of College Science Teaching}\ }\textbf {\bibinfo {volume} {51}},\ \bibinfo {pages} {46} (\bibinfo {year} {2021})}\BibitemShut {NoStop}%
\bibitem [{\citenamefont {Wooten}\ \emph {et~al.}(2018)\citenamefont {Wooten}, \citenamefont {Coble}, \citenamefont {Puckett},\ and\ \citenamefont {Rector}}]{wooten2018investigating}%
  \BibitemOpen
  \bibfield  {author} {\bibinfo {author} {\bibfnamefont {M.~M.}\ \bibnamefont {Wooten}}, \bibinfo {author} {\bibfnamefont {K.}~\bibnamefont {Coble}}, \bibinfo {author} {\bibfnamefont {A.~W.}\ \bibnamefont {Puckett}},\ and\ \bibinfo {author} {\bibfnamefont {T.}~\bibnamefont {Rector}},\ }\bibfield  {title} {\bibinfo {title} {Investigating introductory astronomy students’ perceived impacts from participation in course-based undergraduate research experiences},\ }\href@noop {} {\bibfield  {journal} {\bibinfo  {journal} {Physical Review Physics Education Research}\ }\textbf {\bibinfo {volume} {14}},\ \bibinfo {pages} {010151} (\bibinfo {year} {2018})}\BibitemShut {NoStop}%
\bibitem [{\citenamefont {Rector}\ \emph {et~al.}(2019)\citenamefont {Rector}, \citenamefont {Puckett}, \citenamefont {Wooten}, \citenamefont {Vogt}, \citenamefont {Coble},\ and\ \citenamefont {Pilachowski}}]{rector2019authentic}%
  \BibitemOpen
  \bibfield  {author} {\bibinfo {author} {\bibfnamefont {T.~A.}\ \bibnamefont {Rector}}, \bibinfo {author} {\bibfnamefont {A.~W.}\ \bibnamefont {Puckett}}, \bibinfo {author} {\bibfnamefont {M.~M.}\ \bibnamefont {Wooten}}, \bibinfo {author} {\bibfnamefont {N.~P.}\ \bibnamefont {Vogt}}, \bibinfo {author} {\bibfnamefont {K.}~\bibnamefont {Coble}},\ and\ \bibinfo {author} {\bibfnamefont {C.~A.}\ \bibnamefont {Pilachowski}},\ }\bibfield  {title} {\bibinfo {title} {Authentic research experiences in astronomy to teach the process of science},\ }in\ \href@noop {} {\emph {\bibinfo {booktitle} {Astronomy Education, Volume 1: Evidence-based instruction for introductory courses}}}\ (\bibinfo  {publisher} {IOP Publishing Bristol, UK},\ \bibinfo {year} {2019})\ pp.\ \bibinfo {pages} {7--1}\BibitemShut {NoStop}%
\bibitem [{\citenamefont {Werth}\ \emph {et~al.}(2022)\citenamefont {Werth}, \citenamefont {West},\ and\ \citenamefont {Lewandowski}}]{werth2022impacts}%
  \BibitemOpen
  \bibfield  {author} {\bibinfo {author} {\bibfnamefont {A.}~\bibnamefont {Werth}}, \bibinfo {author} {\bibfnamefont {C.~G.}\ \bibnamefont {West}},\ and\ \bibinfo {author} {\bibfnamefont {H.}~\bibnamefont {Lewandowski}},\ }\bibfield  {title} {\bibinfo {title} {Impacts on student learning, confidence, and affect in a remote, large-enrollment, course-based undergraduate research experience in physics},\ }\href@noop {} {\bibfield  {journal} {\bibinfo  {journal} {Physical Review Physics Education Research}\ }\textbf {\bibinfo {volume} {18}},\ \bibinfo {pages} {010129} (\bibinfo {year} {2022})}\BibitemShut {NoStop}%
\bibitem [{\citenamefont {Oliver}\ \emph {et~al.}(2023)\citenamefont {Oliver}, \citenamefont {Werth},\ and\ \citenamefont {Lewandowski}}]{oliver2023student}%
  \BibitemOpen
  \bibfield  {author} {\bibinfo {author} {\bibfnamefont {K.~A.}\ \bibnamefont {Oliver}}, \bibinfo {author} {\bibfnamefont {A.}~\bibnamefont {Werth}},\ and\ \bibinfo {author} {\bibfnamefont {H.}~\bibnamefont {Lewandowski}},\ }\bibfield  {title} {\bibinfo {title} {Student experiences with authentic research in a remote, introductory course-based undergraduate research experience in physics},\ }\href@noop {} {\bibfield  {journal} {\bibinfo  {journal} {Physical Review Physics Education Research}\ }\textbf {\bibinfo {volume} {19}},\ \bibinfo {pages} {010124} (\bibinfo {year} {2023})}\BibitemShut {NoStop}%
\bibitem [{\citenamefont {Werth}\ \emph {et~al.}(2023)\citenamefont {Werth}, \citenamefont {West}, \citenamefont {Sulaiman},\ and\ \citenamefont {Lewandowski}}]{werth2023enhancing}%
  \BibitemOpen
  \bibfield  {author} {\bibinfo {author} {\bibfnamefont {A.}~\bibnamefont {Werth}}, \bibinfo {author} {\bibfnamefont {C.~G.}\ \bibnamefont {West}}, \bibinfo {author} {\bibfnamefont {N.}~\bibnamefont {Sulaiman}},\ and\ \bibinfo {author} {\bibfnamefont {H.}~\bibnamefont {Lewandowski}},\ }\bibfield  {title} {\bibinfo {title} {Enhancing students’ views of experimental physics through a course-based undergraduate research experience},\ }\href@noop {} {\bibfield  {journal} {\bibinfo  {journal} {Physical Review Physics Education Research}\ }\textbf {\bibinfo {volume} {19}},\ \bibinfo {pages} {020151} (\bibinfo {year} {2023})}\BibitemShut {NoStop}%
\bibitem [{\citenamefont {Hewitt}\ \emph {et~al.}(2023)\citenamefont {Hewitt}, \citenamefont {Simon}, \citenamefont {Mead}, \citenamefont {Grayson}, \citenamefont {Beall}, \citenamefont {Zellem}, \citenamefont {Tock},\ and\ \citenamefont {Pearson}}]{hewitt2023development}%
  \BibitemOpen
  \bibfield  {author} {\bibinfo {author} {\bibfnamefont {H.~B.}\ \bibnamefont {Hewitt}}, \bibinfo {author} {\bibfnamefont {M.~N.}\ \bibnamefont {Simon}}, \bibinfo {author} {\bibfnamefont {C.}~\bibnamefont {Mead}}, \bibinfo {author} {\bibfnamefont {S.}~\bibnamefont {Grayson}}, \bibinfo {author} {\bibfnamefont {G.~L.}\ \bibnamefont {Beall}}, \bibinfo {author} {\bibfnamefont {R.~T.}\ \bibnamefont {Zellem}}, \bibinfo {author} {\bibfnamefont {K.}~\bibnamefont {Tock}},\ and\ \bibinfo {author} {\bibfnamefont {K.~A.}\ \bibnamefont {Pearson}},\ }\bibfield  {title} {\bibinfo {title} {Development and assessment of a course-based undergraduate research experience for online astronomy majors},\ }\href@noop {} {\bibfield  {journal} {\bibinfo  {journal} {Physical Review Physics Education Research}\ }\textbf {\bibinfo {volume} {19}},\ \bibinfo {pages} {020156} (\bibinfo {year} {2023})}\BibitemShut {NoStop}%
\bibitem [{\citenamefont {Duncan}\ and\ \citenamefont {Arthurs}(2012)}]{duncan2012improving}%
  \BibitemOpen
  \bibfield  {author} {\bibinfo {author} {\bibfnamefont {D.}~\bibnamefont {Duncan}}\ and\ \bibinfo {author} {\bibfnamefont {L.}~\bibnamefont {Arthurs}},\ }\bibfield  {title} {\bibinfo {title} {Improving student attitudes about learning science and student scientific reasoning skills},\ }\bibfield  {journal} {\bibinfo  {journal} {Astronomy Education Research}\ }\href {https://doi.org/10.3847/AER2009067} {10.3847/AER2009067} (\bibinfo {year} {2012})\BibitemShut {NoStop}%
\bibitem [{\citenamefont {S{\o}rvik}\ and\ \citenamefont {Mork}(2015)}]{sorvik2015scientific}%
  \BibitemOpen
  \bibfield  {author} {\bibinfo {author} {\bibfnamefont {G.~O.}\ \bibnamefont {S{\o}rvik}}\ and\ \bibinfo {author} {\bibfnamefont {S.~M.}\ \bibnamefont {Mork}},\ }\bibfield  {title} {\bibinfo {title} {Scientific literacy as social practice: Implications for reading and writing in science classrooms},\ }\href {https://doi.org/10.5617/nordina.987} {\bibfield  {journal} {\bibinfo  {journal} {Nordic Studies in Science Education}\ }\textbf {\bibinfo {volume} {11}},\ \bibinfo {pages} {268} (\bibinfo {year} {2015})}\BibitemShut {NoStop}%
\bibitem [{\citenamefont {Szymanski}(2014)}]{szymanski2014instructor}%
  \BibitemOpen
  \bibfield  {author} {\bibinfo {author} {\bibfnamefont {E.~A.}\ \bibnamefont {Szymanski}},\ }\bibfield  {title} {\bibinfo {title} {Instructor feedback in upper-division biology courses: Moving from spelling and syntax to scientific discourse},\ }\bibfield  {journal} {\bibinfo  {journal} {Across the Disciplines}\ }\href {https://doi.org/10.37514/ATD-J.2014.11.2.06} {10.37514/ATD-J.2014.11.2.06} (\bibinfo {year} {2014})\BibitemShut {NoStop}%
\bibitem [{\citenamefont {Pelger}\ and\ \citenamefont {Nilsson}(2016)}]{pelger2016popular}%
  \BibitemOpen
  \bibfield  {author} {\bibinfo {author} {\bibfnamefont {S.}~\bibnamefont {Pelger}}\ and\ \bibinfo {author} {\bibfnamefont {P.}~\bibnamefont {Nilsson}},\ }\bibfield  {title} {\bibinfo {title} {Popular science writing to support students’ learning of science and scientific literacy},\ }\href {https://doi.org/10.1007/s11165-015-9465-y} {\bibfield  {journal} {\bibinfo  {journal} {Research in Science Education}\ }\textbf {\bibinfo {volume} {46}},\ \bibinfo {pages} {439} (\bibinfo {year} {2016})}\BibitemShut {NoStop}%
\bibitem [{\citenamefont {Lewis}\ \emph {et~al.}(2023)\citenamefont {Lewis}, \citenamefont {Supriya}, \citenamefont {Read}, \citenamefont {Dixie}, \citenamefont {Kennison},\ and\ \citenamefont {Friscia}}]{lewis2022effects}%
  \BibitemOpen
  \bibfield  {author} {\bibinfo {author} {\bibfnamefont {B.~L.}\ \bibnamefont {Lewis}}, \bibinfo {author} {\bibfnamefont {K.}~\bibnamefont {Supriya}}, \bibinfo {author} {\bibfnamefont {G.~H.}\ \bibnamefont {Read}}, \bibinfo {author} {\bibfnamefont {K.~L.~I.}\ \bibnamefont {Dixie}}, \bibinfo {author} {\bibfnamefont {R.}~\bibnamefont {Kennison}},\ and\ \bibinfo {author} {\bibfnamefont {A.~R.}\ \bibnamefont {Friscia}},\ }\bibfield  {title} {\bibinfo {title} {Effects of popular science writing instruction on general education student attitudes towards science: A case study in astronomy},\ }\bibfield  {journal} {\bibinfo  {journal} {Astronomy Education Journal}\ }\href {https://doi.org/10.32374/AEJ.2023.3.1.049ra} {10.32374/AEJ.2023.3.1.049ra} (\bibinfo {year} {2023})\BibitemShut {NoStop}%
\bibitem [{\citenamefont {Akbasli}\ \emph {et~al.}(2016)\citenamefont {Akbasli}, \citenamefont {Sahin},\ and\ \citenamefont {Yaykiran}}]{akbasli2016effect}%
  \BibitemOpen
  \bibfield  {author} {\bibinfo {author} {\bibfnamefont {S.}~\bibnamefont {Akbasli}}, \bibinfo {author} {\bibfnamefont {M.}~\bibnamefont {Sahin}},\ and\ \bibinfo {author} {\bibfnamefont {Z.}~\bibnamefont {Yaykiran}},\ }\bibfield  {title} {\bibinfo {title} {The effect of reading comprehension on the performance in science and mathematics.},\ }\href@noop {} {\bibfield  {journal} {\bibinfo  {journal} {Journal of Education and Practice}\ }\textbf {\bibinfo {volume} {7}},\ \bibinfo {pages} {108} (\bibinfo {year} {2016})}\BibitemShut {NoStop}%
\bibitem [{\citenamefont {Heron}\ \emph {et~al.}(2019)\citenamefont {Heron}, \citenamefont {McNeil},\ and\ \citenamefont {Co-chairs}}]{heron2019joint}%
  \BibitemOpen
  \bibfield  {author} {\bibinfo {author} {\bibfnamefont {H.}~\bibnamefont {Heron}}, \bibinfo {author} {\bibfnamefont {L.}~\bibnamefont {McNeil}},\ and\ \bibinfo {author} {\bibnamefont {Co-chairs}},\ }\href@noop {} {\bibinfo {title} {Joint task force on undergraduate physics programs. phys21: preparing physics student for 21st-century careers}} (\bibinfo {year} {2019})\BibitemShut {NoStop}%
\bibitem [{\citenamefont {Williams}(2020)}]{williams2020first}%
  \BibitemOpen
  \bibfield  {author} {\bibinfo {author} {\bibfnamefont {D.}~\bibnamefont {Williams}},\ }\bibfield  {title} {\bibinfo {title} {How first-year comp can save the world},\ }\href@noop {} {\bibfield  {journal} {\bibinfo  {journal} {""}\ } (\bibinfo {year} {2020})}\BibitemShut {NoStop}%
\bibitem [{\citenamefont {Van~Schaik}\ \emph {et~al.}(2018)\citenamefont {Van~Schaik}, \citenamefont {Volman}, \citenamefont {Admiraal},\ and\ \citenamefont {Schenke}}]{van2018barriers}%
  \BibitemOpen
  \bibfield  {author} {\bibinfo {author} {\bibfnamefont {P.}~\bibnamefont {Van~Schaik}}, \bibinfo {author} {\bibfnamefont {M.}~\bibnamefont {Volman}}, \bibinfo {author} {\bibfnamefont {W.}~\bibnamefont {Admiraal}},\ and\ \bibinfo {author} {\bibfnamefont {W.}~\bibnamefont {Schenke}},\ }\bibfield  {title} {\bibinfo {title} {Barriers and conditions for teachers’ utilisation of academic knowledge},\ }\href@noop {} {\bibfield  {journal} {\bibinfo  {journal} {International Journal of Educational Research}\ }\textbf {\bibinfo {volume} {90}},\ \bibinfo {pages} {50} (\bibinfo {year} {2018})}\BibitemShut {NoStop}%
\bibitem [{\citenamefont {Shkedi}(1998)}]{shkedi1998teachers}%
  \BibitemOpen
  \bibfield  {author} {\bibinfo {author} {\bibfnamefont {A.}~\bibnamefont {Shkedi}},\ }\bibfield  {title} {\bibinfo {title} {Teachers' attitudes towards research: A challenge for qualitative researchers},\ }\href@noop {} {\bibfield  {journal} {\bibinfo  {journal} {International Journal of Qualitative Studies in Education}\ }\textbf {\bibinfo {volume} {11}},\ \bibinfo {pages} {559} (\bibinfo {year} {1998})}\BibitemShut {NoStop}%
\bibitem [{\citenamefont {Young}\ \emph {et~al.}(2022)\citenamefont {Young}, \citenamefont {Lewis}, \citenamefont {Kerr},\ and\ \citenamefont {Nair}}]{young2022using}%
  \BibitemOpen
  \bibfield  {author} {\bibinfo {author} {\bibfnamefont {N.~T.}\ \bibnamefont {Young}}, \bibinfo {author} {\bibfnamefont {B.~L.}\ \bibnamefont {Lewis}}, \bibinfo {author} {\bibfnamefont {E.}~\bibnamefont {Kerr}},\ and\ \bibinfo {author} {\bibfnamefont {P.~H.}\ \bibnamefont {Nair}},\ }\bibfield  {title} {\bibinfo {title} {Using blogs to make peer-reviewed research more accessible},\ }\href@noop {} {\bibfield  {journal} {\bibinfo  {journal} {arXiv preprint arXiv:2209.07627}\ } (\bibinfo {year} {2022})}\BibitemShut {NoStop}%
\bibitem [{\citenamefont {Adler-Kassner}\ and\ \citenamefont {Estrem}(2007)}]{adler2007reading}%
  \BibitemOpen
  \bibfield  {author} {\bibinfo {author} {\bibfnamefont {L.}~\bibnamefont {Adler-Kassner}}\ and\ \bibinfo {author} {\bibfnamefont {H.}~\bibnamefont {Estrem}},\ }\bibfield  {title} {\bibinfo {title} {Reading practices in the writing classroom},\ }\href@noop {} {\bibfield  {journal} {\bibinfo  {journal} {Writing Program Administration}\ }\textbf {\bibinfo {volume} {31}},\ \bibinfo {pages} {35} (\bibinfo {year} {2007})}\BibitemShut {NoStop}%
\bibitem [{\citenamefont {Henderson}\ and\ \citenamefont {Dancy}(2007)}]{henderson2007barriers}%
  \BibitemOpen
  \bibfield  {author} {\bibinfo {author} {\bibfnamefont {C.}~\bibnamefont {Henderson}}\ and\ \bibinfo {author} {\bibfnamefont {M.~H.}\ \bibnamefont {Dancy}},\ }\bibfield  {title} {\bibinfo {title} {Barriers to the use of research-based instructional strategies: The influence of both individual and situational characteristics},\ }\href@noop {} {\bibfield  {journal} {\bibinfo  {journal} {Physical Review Special Topics-Physics Education Research}\ }\textbf {\bibinfo {volume} {3}},\ \bibinfo {pages} {020102} (\bibinfo {year} {2007})}\BibitemShut {NoStop}%
\bibitem [{\citenamefont {Dancy}\ and\ \citenamefont {Henderson}(2010)}]{dancy2010pedagogical}%
  \BibitemOpen
  \bibfield  {author} {\bibinfo {author} {\bibfnamefont {M.}~\bibnamefont {Dancy}}\ and\ \bibinfo {author} {\bibfnamefont {C.}~\bibnamefont {Henderson}},\ }\bibfield  {title} {\bibinfo {title} {Pedagogical practices and instructional change of physics faculty},\ }\href@noop {} {\bibfield  {journal} {\bibinfo  {journal} {American Journal of Physics}\ }\textbf {\bibinfo {volume} {78}},\ \bibinfo {pages} {1056} (\bibinfo {year} {2010})}\BibitemShut {NoStop}%
\bibitem [{\citenamefont {Koch}\ and\ \citenamefont {Eckstein}(1991)}]{koch1991improvement}%
  \BibitemOpen
  \bibfield  {author} {\bibinfo {author} {\bibfnamefont {A.}~\bibnamefont {Koch}}\ and\ \bibinfo {author} {\bibfnamefont {S.~G.}\ \bibnamefont {Eckstein}},\ }\bibfield  {title} {\bibinfo {title} {Improvement of reading comprehension of physics texts by students’ question formulation},\ }\href@noop {} {\bibfield  {journal} {\bibinfo  {journal} {International Journal of Science Education}\ }\textbf {\bibinfo {volume} {13}},\ \bibinfo {pages} {473} (\bibinfo {year} {1991})}\BibitemShut {NoStop}%
\bibitem [{\citenamefont {Koch}(2001)}]{koch2001training}%
  \BibitemOpen
  \bibfield  {author} {\bibinfo {author} {\bibfnamefont {A.}~\bibnamefont {Koch}},\ }\bibfield  {title} {\bibinfo {title} {Training in metacognition and comprehension of physics texts},\ }\href@noop {} {\bibfield  {journal} {\bibinfo  {journal} {Science education}\ }\textbf {\bibinfo {volume} {85}},\ \bibinfo {pages} {758} (\bibinfo {year} {2001})}\BibitemShut {NoStop}%
\bibitem [{\citenamefont {Garland}\ and\ \citenamefont {Ratay}(2007)}]{garland2007using}%
  \BibitemOpen
  \bibfield  {author} {\bibinfo {author} {\bibfnamefont {C.~A.}\ \bibnamefont {Garland}}\ and\ \bibinfo {author} {\bibfnamefont {D.~L.}\ \bibnamefont {Ratay}},\ }\bibfield  {title} {\bibinfo {title} {Using literacy techniques to teach astronomy to non-science majors.},\ }\href@noop {} {\bibfield  {journal} {\bibinfo  {journal} {Astronomy Education Review}\ }\textbf {\bibinfo {volume} {6}} (\bibinfo {year} {2007})}\BibitemShut {NoStop}%
\bibitem [{\citenamefont {Baram-Tsabari}\ and\ \citenamefont {Yarden}(2011)}]{baram2011quantifying}%
  \BibitemOpen
  \bibfield  {author} {\bibinfo {author} {\bibfnamefont {A.}~\bibnamefont {Baram-Tsabari}}\ and\ \bibinfo {author} {\bibfnamefont {A.}~\bibnamefont {Yarden}},\ }\bibfield  {title} {\bibinfo {title} {Quantifying the gender gap in science interests},\ }\href@noop {} {\bibfield  {journal} {\bibinfo  {journal} {International Journal of Science and Mathematics Education}\ }\textbf {\bibinfo {volume} {9}},\ \bibinfo {pages} {523} (\bibinfo {year} {2011})}\BibitemShut {NoStop}%
\bibitem [{\citenamefont {Susiati}\ \emph {et~al.}(2018)\citenamefont {Susiati}, \citenamefont {Adisyahputra},\ and\ \citenamefont {Miarsyah}}]{susiati2018correlation}%
  \BibitemOpen
  \bibfield  {author} {\bibinfo {author} {\bibfnamefont {A.}~\bibnamefont {Susiati}}, \bibinfo {author} {\bibfnamefont {A.}~\bibnamefont {Adisyahputra}},\ and\ \bibinfo {author} {\bibfnamefont {M.}~\bibnamefont {Miarsyah}},\ }\bibfield  {title} {\bibinfo {title} {Correlation of comprehension reading skill and higher-order thinking skill with scientific literacy skill of senior high school biology teacher},\ }\href@noop {} {\bibfield  {journal} {\bibinfo  {journal} {Biosfer: Jurnal Pendidikan Biologi}\ }\textbf {\bibinfo {volume} {11}},\ \bibinfo {pages} {1} (\bibinfo {year} {2018})}\BibitemShut {NoStop}%
\bibitem [{\citenamefont {Kohler}\ \emph {et~al.}(2018)\citenamefont {Kohler}, \citenamefont {Collaboration} \emph {et~al.}}]{kohler2018aas}%
  \BibitemOpen
  \bibfield  {author} {\bibinfo {author} {\bibfnamefont {S.}~\bibnamefont {Kohler}}, \bibinfo {author} {\bibfnamefont {A.}~\bibnamefont {Collaboration}}, \emph {et~al.},\ }\bibfield  {title} {\bibinfo {title} {Aas nova and astrobites as bridges between astronomy communities},\ }\href@noop {} {\bibfield  {journal} {\bibinfo  {journal} {Proceedings of the International Astronomical Union}\ }\textbf {\bibinfo {volume} {14}},\ \bibinfo {pages} {524} (\bibinfo {year} {2018})}\BibitemShut {NoStop}%
\bibitem [{\citenamefont {{Sanders}}\ \emph {et~al.}(2012)\citenamefont {{Sanders}}, \citenamefont {{Kohler}}, \citenamefont {{Newton}},\ and\ \citenamefont {{Astrobites Collaboration}}}]{sanders2012preparing}%
  \BibitemOpen
  \bibfield  {author} {\bibinfo {author} {\bibfnamefont {N.~E.}\ \bibnamefont {{Sanders}}}, \bibinfo {author} {\bibfnamefont {S.}~\bibnamefont {{Kohler}}}, \bibinfo {author} {\bibfnamefont {E.}~\bibnamefont {{Newton}}},\ and\ \bibinfo {author} {\bibnamefont {{Astrobites Collaboration}}},\ }\bibfield  {title} {\bibinfo {title} {{Preparing Undergraduates for Research Careers: Using Astrobites in the Classroom}},\ }\href {https://doi.org/10.3847/AER2012030} {\bibfield  {journal} {\bibinfo  {journal} {Astronomy Education Review}\ }\textbf {\bibinfo {volume} {11}},\ \bibinfo {pages} {010201} (\bibinfo {year} {2012})},\ \Eprint {https://arxiv.org/abs/1208.4791} {arXiv:1208.4791 [physics.ed-ph]} \BibitemShut {NoStop}%
\bibitem [{\citenamefont {{Sanders}}\ \emph {et~al.}(2017)\citenamefont {{Sanders}}, \citenamefont {{Kohler}}, \citenamefont {{Faesi}}, \citenamefont {{Villar}},\ and\ \citenamefont {{Zevin}}}]{sanders2017incorporating}%
  \BibitemOpen
  \bibfield  {author} {\bibinfo {author} {\bibfnamefont {N.~E.}\ \bibnamefont {{Sanders}}}, \bibinfo {author} {\bibfnamefont {S.}~\bibnamefont {{Kohler}}}, \bibinfo {author} {\bibfnamefont {C.}~\bibnamefont {{Faesi}}}, \bibinfo {author} {\bibfnamefont {A.}~\bibnamefont {{Villar}}},\ and\ \bibinfo {author} {\bibfnamefont {M.}~\bibnamefont {{Zevin}}},\ }\bibfield  {title} {\bibinfo {title} {{Incorporating current research into formal higher education settings using Astrobites}},\ }\href {https://doi.org/10.1119/1.4991506} {\bibfield  {journal} {\bibinfo  {journal} {American Journal of Physics}\ }\textbf {\bibinfo {volume} {85}},\ \bibinfo {pages} {741} (\bibinfo {year} {2017})},\ \Eprint {https://arxiv.org/abs/1706.01165} {arXiv:1706.01165 [physics.ed-ph]} \BibitemShut {NoStop}%
\bibitem [{\citenamefont {Bartlett}\ \emph {et~al.}(2018)\citenamefont {Bartlett}, \citenamefont {Fitzgerald}, \citenamefont {McKinnon}, \citenamefont {Danaia}, \citenamefont {Lazendic-Galloway} \emph {et~al.}}]{bartlett2018astronomy}%
  \BibitemOpen
  \bibfield  {author} {\bibinfo {author} {\bibfnamefont {S.}~\bibnamefont {Bartlett}}, \bibinfo {author} {\bibfnamefont {M.~T.}\ \bibnamefont {Fitzgerald}}, \bibinfo {author} {\bibfnamefont {D.~H.}\ \bibnamefont {McKinnon}}, \bibinfo {author} {\bibfnamefont {L.}~\bibnamefont {Danaia}}, \bibinfo {author} {\bibfnamefont {J.}~\bibnamefont {Lazendic-Galloway}}, \emph {et~al.},\ }\bibfield  {title} {\bibinfo {title} {Astronomy and science student attitudes (assa): a short review and validation of a new instrument},\ }\href {https://doi.org/10.19030/jaese.v5i1.10190} {\bibfield  {journal} {\bibinfo  {journal} {Journal of Astronomy \& Earth Sciences Education (JAESE)}\ }\textbf {\bibinfo {volume} {5}},\ \bibinfo {pages} {1} (\bibinfo {year} {2018})}\BibitemShut {NoStop}%
\bibitem [{\citenamefont {Elby}(2001)}]{elby2001helping}%
  \BibitemOpen
  \bibfield  {author} {\bibinfo {author} {\bibfnamefont {A.}~\bibnamefont {Elby}},\ }\bibfield  {title} {\bibinfo {title} {Helping physics students learn how to learn},\ }\href {https://doi.org/10.1119/1.1377283} {\bibfield  {journal} {\bibinfo  {journal} {American Journal of Physics}\ }\textbf {\bibinfo {volume} {69}},\ \bibinfo {pages} {S54} (\bibinfo {year} {2001})}\BibitemShut {NoStop}%
\bibitem [{\citenamefont {Estrada}\ \emph {et~al.}(2011)\citenamefont {Estrada}, \citenamefont {Woodcock}, \citenamefont {Hernandez},\ and\ \citenamefont {Schultz}}]{estrada2011toward}%
  \BibitemOpen
  \bibfield  {author} {\bibinfo {author} {\bibfnamefont {M.}~\bibnamefont {Estrada}}, \bibinfo {author} {\bibfnamefont {A.}~\bibnamefont {Woodcock}}, \bibinfo {author} {\bibfnamefont {P.~R.}\ \bibnamefont {Hernandez}},\ and\ \bibinfo {author} {\bibfnamefont {P.}~\bibnamefont {Schultz}},\ }\bibfield  {title} {\bibinfo {title} {Toward a model of social influence that explains minority student integration into the scientific community.},\ }\href {https://doi.org/10.1037/a0020743} {\bibfield  {journal} {\bibinfo  {journal} {Journal of educational psychology}\ }\textbf {\bibinfo {volume} {103}},\ \bibinfo {pages} {206} (\bibinfo {year} {2011})}\BibitemShut {NoStop}%
\bibitem [{\citenamefont {Espinosa}\ \emph {et~al.}(2019)\citenamefont {Espinosa}, \citenamefont {Miller}, \citenamefont {Araujo},\ and\ \citenamefont {Mazur}}]{espinosa2019reducing}%
  \BibitemOpen
  \bibfield  {author} {\bibinfo {author} {\bibfnamefont {T.}~\bibnamefont {Espinosa}}, \bibinfo {author} {\bibfnamefont {K.}~\bibnamefont {Miller}}, \bibinfo {author} {\bibfnamefont {I.}~\bibnamefont {Araujo}},\ and\ \bibinfo {author} {\bibfnamefont {E.}~\bibnamefont {Mazur}},\ }\bibfield  {title} {\bibinfo {title} {Reducing the gender gap in students' physics self-efficacy in a team- and project-based introductory physics class},\ }\href {https://doi.org/10.1103/PhysRevPhysEducRes.15.010132} {\bibfield  {journal} {\bibinfo  {journal} {Phys. Rev. Phys. Educ. Res.}\ }\textbf {\bibinfo {volume} {15}},\ \bibinfo {pages} {010132} (\bibinfo {year} {2019})}\BibitemShut {NoStop}%
\bibitem [{\citenamefont {Litwin}\ and\ \citenamefont {Fink}(1995)}]{litwin1995measure}%
  \BibitemOpen
  \bibfield  {author} {\bibinfo {author} {\bibfnamefont {M.~S.}\ \bibnamefont {Litwin}}\ and\ \bibinfo {author} {\bibfnamefont {A.}~\bibnamefont {Fink}},\ }\href@noop {} {\emph {\bibinfo {title} {How to measure survey reliability and validity}}},\ Vol.~\bibinfo {volume} {7}\ (\bibinfo  {publisher} {Sage},\ \bibinfo {year} {1995})\BibitemShut {NoStop}%
\bibitem [{\citenamefont {Taherdoost}(2016)}]{taherdoost2016validity}%
  \BibitemOpen
  \bibfield  {author} {\bibinfo {author} {\bibfnamefont {H.}~\bibnamefont {Taherdoost}},\ }\bibfield  {title} {\bibinfo {title} {Validity and reliability of the research instrument; how to test the validation of a questionnaire/survey in a research},\ }\bibfield  {journal} {\bibinfo  {journal} {How to test the validation of a questionnaire/survey in a research (August 10, 2016)}\ }\href {https://doi.org/10.2139/ssrn.3205040} {10.2139/ssrn.3205040} (\bibinfo {year} {2016})\BibitemShut {NoStop}%
\bibitem [{\citenamefont {Pepper}\ \emph {et~al.}(2018)\citenamefont {Pepper}, \citenamefont {Hodgen}, \citenamefont {Lamesoo}, \citenamefont {K{\~o}iv},\ and\ \citenamefont {Tolboom}}]{pepper2018think}%
  \BibitemOpen
  \bibfield  {author} {\bibinfo {author} {\bibfnamefont {D.}~\bibnamefont {Pepper}}, \bibinfo {author} {\bibfnamefont {J.}~\bibnamefont {Hodgen}}, \bibinfo {author} {\bibfnamefont {K.}~\bibnamefont {Lamesoo}}, \bibinfo {author} {\bibfnamefont {P.}~\bibnamefont {K{\~o}iv}},\ and\ \bibinfo {author} {\bibfnamefont {J.}~\bibnamefont {Tolboom}},\ }\bibfield  {title} {\bibinfo {title} {Think aloud: using cognitive interviewing to validate the pisa assessment of student self-efficacy in mathematics},\ }\href@noop {} {\bibfield  {journal} {\bibinfo  {journal} {International Journal of Research \& Method in Education}\ }\textbf {\bibinfo {volume} {41}},\ \bibinfo {pages} {3} (\bibinfo {year} {2018})}\BibitemShut {NoStop}%
\bibitem [{\citenamefont {Wolcott}\ and\ \citenamefont {Lobczowski}(2021)}]{wolcott2021using}%
  \BibitemOpen
  \bibfield  {author} {\bibinfo {author} {\bibfnamefont {M.~D.}\ \bibnamefont {Wolcott}}\ and\ \bibinfo {author} {\bibfnamefont {N.~G.}\ \bibnamefont {Lobczowski}},\ }\bibfield  {title} {\bibinfo {title} {Using cognitive interviews and think-aloud protocols to understand thought processes},\ }\href@noop {} {\bibfield  {journal} {\bibinfo  {journal} {Currents in Pharmacy Teaching and Learning}\ }\textbf {\bibinfo {volume} {13}},\ \bibinfo {pages} {181} (\bibinfo {year} {2021})}\BibitemShut {NoStop}%
\bibitem [{\citenamefont {Ponce}\ and\ \citenamefont {Pag{\'a}n-Maldonado}(2015)}]{ponce2015mixed}%
  \BibitemOpen
  \bibfield  {author} {\bibinfo {author} {\bibfnamefont {O.~A.}\ \bibnamefont {Ponce}}\ and\ \bibinfo {author} {\bibfnamefont {N.}~\bibnamefont {Pag{\'a}n-Maldonado}},\ }\bibfield  {title} {\bibinfo {title} {Mixed methods research in education: Capturing the complexity of the profession},\ }\href {https://doi.org/10.18562/ijee.2015.0005} {\bibfield  {journal} {\bibinfo  {journal} {International journal of educational excellence}\ }\textbf {\bibinfo {volume} {1}},\ \bibinfo {pages} {111} (\bibinfo {year} {2015})}\BibitemShut {NoStop}%
\bibitem [{\citenamefont {Formanek}\ \emph {et~al.}(2019)\citenamefont {Formanek}, \citenamefont {Buxner}, \citenamefont {Impey},\ and\ \citenamefont {Wenger}}]{formanek2019}%
  \BibitemOpen
  \bibfield  {author} {\bibinfo {author} {\bibfnamefont {M.}~\bibnamefont {Formanek}}, \bibinfo {author} {\bibfnamefont {S.}~\bibnamefont {Buxner}}, \bibinfo {author} {\bibfnamefont {C.}~\bibnamefont {Impey}},\ and\ \bibinfo {author} {\bibfnamefont {M.}~\bibnamefont {Wenger}},\ }\bibfield  {title} {\bibinfo {title} {Relationship between learners' motivation and course engagement in an astronomy massive open online course},\ }\href {https://doi.org/10.1103/PhysRevPhysEducRes.15.020140} {\bibfield  {journal} {\bibinfo  {journal} {Phys. Rev. Phys. Educ. Res.}\ }\textbf {\bibinfo {volume} {15}},\ \bibinfo {pages} {020140} (\bibinfo {year} {2019})}\BibitemShut {NoStop}%
\bibitem [{\citenamefont {Zou}\ \emph {et~al.}(2024)\citenamefont {Zou}, \citenamefont {Xue}, \citenamefont {Jin}, \citenamefont {Huang},\ and\ \citenamefont {Li}}]{PhysRevPhysEducRes.20.020107}%
  \BibitemOpen
  \bibfield  {author} {\bibinfo {author} {\bibfnamefont {Y.}~\bibnamefont {Zou}}, \bibinfo {author} {\bibfnamefont {X.}~\bibnamefont {Xue}}, \bibinfo {author} {\bibfnamefont {L.}~\bibnamefont {Jin}}, \bibinfo {author} {\bibfnamefont {X.}~\bibnamefont {Huang}},\ and\ \bibinfo {author} {\bibfnamefont {Y.}~\bibnamefont {Li}},\ }\bibfield  {title} {\bibinfo {title} {Assessment of conceptual understanding in student learning of evaporation},\ }\href {https://doi.org/10.1103/PhysRevPhysEducRes.20.020107} {\bibfield  {journal} {\bibinfo  {journal} {Phys. Rev. Phys. Educ. Res.}\ }\textbf {\bibinfo {volume} {20}},\ \bibinfo {pages} {020107} (\bibinfo {year} {2024})}\BibitemShut {NoStop}%
\bibitem [{\citenamefont {Ding}\ \emph {et~al.}(2024)\citenamefont {Ding}, \citenamefont {Zhu}, \citenamefont {Bian},\ and\ \citenamefont {Bao}}]{PhysRevPhysEducRes.20.020141}%
  \BibitemOpen
  \bibfield  {author} {\bibinfo {author} {\bibfnamefont {Y.}~\bibnamefont {Ding}}, \bibinfo {author} {\bibfnamefont {G.}~\bibnamefont {Zhu}}, \bibinfo {author} {\bibfnamefont {Q.}~\bibnamefont {Bian}},\ and\ \bibinfo {author} {\bibfnamefont {L.}~\bibnamefont {Bao}},\ }\bibfield  {title} {\bibinfo {title} {Analysis of students' conceptual change in learning newton's third law with an integrated framework of model analysis and knowledge integration},\ }\href {https://doi.org/10.1103/PhysRevPhysEducRes.20.020141} {\bibfield  {journal} {\bibinfo  {journal} {Phys. Rev. Phys. Educ. Res.}\ }\textbf {\bibinfo {volume} {20}},\ \bibinfo {pages} {020141} (\bibinfo {year} {2024})}\BibitemShut {NoStop}%
\bibitem [{\citenamefont {Cronbach}(1951)}]{Cronbach1952}%
  \BibitemOpen
  \bibfield  {author} {\bibinfo {author} {\bibfnamefont {L.~J.}\ \bibnamefont {Cronbach}},\ }\bibfield  {title} {\bibinfo {title} {Coefficient alpha and the internal structure of tests},\ }\href {https://link.springer.com/article/10.1007/BF02310555#citeas} {\bibfield  {journal} {\bibinfo  {journal} {Psychometrika}\ }\textbf {\bibinfo {volume} {16}},\ \bibinfo {pages} {297} (\bibinfo {year} {1951})}\BibitemShut {NoStop}%
\bibitem [{AIP(2024{\natexlab{a}})}]{AIP_2024a}%
  \BibitemOpen
  \href {https://ww2.aip.org/statistics/roster-of-physics-department-with-enrollment-and-degree-data-2023} {} (\bibinfo {year} {2024}{\natexlab{a}})\BibitemShut {NoStop}%
\bibitem [{AIP(2024{\natexlab{b}})}]{AIP_2024b}%
  \BibitemOpen
  \href {https://ww2.aip.org/statistics/roster-of-astronomy-departments-with-enrollment-and-degree-data-2023} {} (\bibinfo {year} {2024}{\natexlab{b}})\BibitemShut {NoStop}%
\bibitem [{\citenamefont {Zhai}\ \emph {et~al.}(2010)\citenamefont {Zhai}, \citenamefont {Raver}, \citenamefont {Jones}, \citenamefont {Li-Grining}, \citenamefont {Pressler},\ and\ \citenamefont {Gao}}]{zhai2010dosage}%
  \BibitemOpen
  \bibfield  {author} {\bibinfo {author} {\bibfnamefont {F.}~\bibnamefont {Zhai}}, \bibinfo {author} {\bibfnamefont {C.~C.}\ \bibnamefont {Raver}}, \bibinfo {author} {\bibfnamefont {S.~M.}\ \bibnamefont {Jones}}, \bibinfo {author} {\bibfnamefont {C.~P.}\ \bibnamefont {Li-Grining}}, \bibinfo {author} {\bibfnamefont {E.}~\bibnamefont {Pressler}},\ and\ \bibinfo {author} {\bibfnamefont {Q.}~\bibnamefont {Gao}},\ }\bibfield  {title} {\bibinfo {title} {Dosage effects on school readiness: Evidence from a randomized classroom-based intervention},\ }\href@noop {} {\bibfield  {journal} {\bibinfo  {journal} {Social Service Review}\ }\textbf {\bibinfo {volume} {84}},\ \bibinfo {pages} {615} (\bibinfo {year} {2010})}\BibitemShut {NoStop}%
\bibitem [{\citenamefont {Vallat}(2018)}]{Vallat2018}%
  \BibitemOpen
  \bibfield  {author} {\bibinfo {author} {\bibfnamefont {R.}~\bibnamefont {Vallat}},\ }\bibfield  {title} {\bibinfo {title} {Pingouin: statistics in python},\ }\href {https://doi.org/10.21105/joss.01026} {\bibfield  {journal} {\bibinfo  {journal} {Journal of Open Source Software}\ }\textbf {\bibinfo {volume} {3}},\ \bibinfo {pages} {1026} (\bibinfo {year} {2018})}\BibitemShut {NoStop}%
\bibitem [{\citenamefont {Bakan}(1966)}]{bakan1966test}%
  \BibitemOpen
  \bibfield  {author} {\bibinfo {author} {\bibfnamefont {D.}~\bibnamefont {Bakan}},\ }\bibfield  {title} {\bibinfo {title} {The test of significance in psychological research.},\ }\href {https://doi.org/10.1037/h0020412} {\bibfield  {journal} {\bibinfo  {journal} {Psychological bulletin}\ }\textbf {\bibinfo {volume} {66}},\ \bibinfo {pages} {423} (\bibinfo {year} {1966})}\BibitemShut {NoStop}%
\bibitem [{\citenamefont {Bartoš}\ and\ \citenamefont {Wagenmakers}(2023)}]{bartos2023general}%
  \BibitemOpen
  \bibfield  {author} {\bibinfo {author} {\bibfnamefont {F.}~\bibnamefont {Bartoš}}\ and\ \bibinfo {author} {\bibfnamefont {E.-J.}\ \bibnamefont {Wagenmakers}},\ }\bibfield  {title} {\bibinfo {title} {A general approximation to nested bayes factors with informed priors},\ }\href {https://doi.org/https://doi.org/10.1002/sta4.600} {\bibfield  {journal} {\bibinfo  {journal} {Stat}\ }\textbf {\bibinfo {volume} {12}},\ \bibinfo {pages} {e600} (\bibinfo {year} {2023})},\ \Eprint {https://arxiv.org/abs/https://onlinelibrary.wiley.com/doi/pdf/10.1002/sta4.600} {https://onlinelibrary.wiley.com/doi/pdf/10.1002/sta4.600} \BibitemShut {NoStop}%
\bibitem [{\citenamefont {Lee}\ and\ \citenamefont {Wagenmakers}(2014)}]{lee2014bayesian}%
  \BibitemOpen
  \bibfield  {author} {\bibinfo {author} {\bibfnamefont {M.~D.}\ \bibnamefont {Lee}}\ and\ \bibinfo {author} {\bibfnamefont {E.-J.}\ \bibnamefont {Wagenmakers}},\ }\href@noop {} {\emph {\bibinfo {title} {Bayesian cognitive modeling: A practical course}}}\ (\bibinfo  {publisher} {Cambridge university press},\ \bibinfo {year} {2014})\BibitemShut {NoStop}%
\bibitem [{\citenamefont {Clarke}\ \emph {et~al.}(2015)\citenamefont {Clarke}, \citenamefont {Braun},\ and\ \citenamefont {Hayfield}}]{clarke2015thematic}%
  \BibitemOpen
  \bibfield  {author} {\bibinfo {author} {\bibfnamefont {V.}~\bibnamefont {Clarke}}, \bibinfo {author} {\bibfnamefont {V.}~\bibnamefont {Braun}},\ and\ \bibinfo {author} {\bibfnamefont {N.}~\bibnamefont {Hayfield}},\ }\bibfield  {title} {\bibinfo {title} {Thematic analysis},\ }\href {https://doi.org/10.1007/978-981-10-5251-4_103} {\bibfield  {journal} {\bibinfo  {journal} {Qualitative psychology: A practical guide to research methods}\ }\textbf {\bibinfo {volume} {3}},\ \bibinfo {pages} {222} (\bibinfo {year} {2015})}\BibitemShut {NoStop}%
\bibitem [{\citenamefont {Gisev}\ \emph {et~al.}(2013)\citenamefont {Gisev}, \citenamefont {Bell},\ and\ \citenamefont {Chen}}]{GISEV2013330}%
  \BibitemOpen
  \bibfield  {author} {\bibinfo {author} {\bibfnamefont {N.}~\bibnamefont {Gisev}}, \bibinfo {author} {\bibfnamefont {J.~S.}\ \bibnamefont {Bell}},\ and\ \bibinfo {author} {\bibfnamefont {T.~F.}\ \bibnamefont {Chen}},\ }\bibfield  {title} {\bibinfo {title} {Interrater agreement and interrater reliability: Key concepts, approaches, and applications},\ }\href {https://doi.org/https://doi.org/10.1016/j.sapharm.2012.04.004} {\bibfield  {journal} {\bibinfo  {journal} {Research in Social and Administrative Pharmacy}\ }\textbf {\bibinfo {volume} {9}},\ \bibinfo {pages} {330} (\bibinfo {year} {2013})}\BibitemShut {NoStop}%
\bibitem [{\citenamefont {Grayson}\ and\ \citenamefont {Rust}(2001)}]{Grayson2001}%
  \BibitemOpen
  \bibfield  {author} {\bibinfo {author} {\bibfnamefont {K.}~\bibnamefont {Grayson}}\ and\ \bibinfo {author} {\bibfnamefont {R.}~\bibnamefont {Rust}},\ }\bibfield  {title} {{\selectlanguage {English (US)}\bibinfo {title} {Interrater reliability}},\ }\href {https://doi.org/10.1207/15327660151043998} {\bibfield  {journal} {\bibinfo  {journal} {Journal of Consumer Psychology}\ }\textbf {\bibinfo {volume} {10}},\ \bibinfo {pages} {71} (\bibinfo {year} {2001})}\BibitemShut {NoStop}%
\bibitem [{\citenamefont {Campbell}\ \emph {et~al.}(2013)\citenamefont {Campbell}, \citenamefont {Quincy}, \citenamefont {Osserman},\ and\ \citenamefont {Pedersen}}]{Campbell_2013}%
  \BibitemOpen
  \bibfield  {author} {\bibinfo {author} {\bibfnamefont {J.~L.}\ \bibnamefont {Campbell}}, \bibinfo {author} {\bibfnamefont {C.}~\bibnamefont {Quincy}}, \bibinfo {author} {\bibfnamefont {J.}~\bibnamefont {Osserman}},\ and\ \bibinfo {author} {\bibfnamefont {O.~K.}\ \bibnamefont {Pedersen}},\ }\bibfield  {title} {\bibinfo {title} {Coding in-depth semistructured interviews: Problems of unitization and intercoder reliability and agreement},\ }\href {https://doi.org/10.1177/0049124113500475} {\bibfield  {journal} {\bibinfo  {journal} {Sociological Methods \& Research}\ }\textbf {\bibinfo {volume} {42}},\ \bibinfo {pages} {294} (\bibinfo {year} {2013})},\ \Eprint {https://arxiv.org/abs/https://doi.org/10.1177/0049124113500475} {https://doi.org/10.1177/0049124113500475} \BibitemShut {NoStop}%
\bibitem [{\citenamefont {Stemler}(2004)}]{Stemler2004ACO}%
  \BibitemOpen
  \bibfield  {author} {\bibinfo {author} {\bibfnamefont {S.~E.}\ \bibnamefont {Stemler}},\ }\bibfield  {title} {\bibinfo {title} {A comparison of consensus, consistency, and measurement approaches to estimating interrater reliability},\ }\href {https://api.semanticscholar.org/CorpusID:8705056} {\bibfield  {journal} {\bibinfo  {journal} {Practical Assessment, Research and Evaluation}\ }\textbf {\bibinfo {volume} {9}},\ \bibinfo {pages} {4} (\bibinfo {year} {2004})}\BibitemShut {NoStop}%
\bibitem [{\citenamefont {Kurasaki}(2000)}]{kurasaki}%
  \BibitemOpen
  \bibfield  {author} {\bibinfo {author} {\bibfnamefont {K.~S.}\ \bibnamefont {Kurasaki}},\ }\bibfield  {title} {\bibinfo {title} {Intercoder reliability for validating conclusions drawn from open-ended interview data},\ }\href {https://doi.org/10.1177/1525822X0001200301} {\bibfield  {journal} {\bibinfo  {journal} {Field Methods}\ }\textbf {\bibinfo {volume} {12}},\ \bibinfo {pages} {179} (\bibinfo {year} {2000})},\ \Eprint {https://arxiv.org/abs/https://doi.org/10.1177/1525822X0001200301} {https://doi.org/10.1177/1525822X0001200301} \BibitemShut {NoStop}%
\bibitem [{\citenamefont {Holdford}(2008)}]{HOLDFORD2008173}%
  \BibitemOpen
  \bibfield  {author} {\bibinfo {author} {\bibfnamefont {D.}~\bibnamefont {Holdford}},\ }\bibfield  {title} {\bibinfo {title} {Content analysis methods for conducting research in social and administrative pharmacy},\ }\href {https://doi.org/https://doi.org/10.1016/j.sapharm.2007.03.003} {\bibfield  {journal} {\bibinfo  {journal} {Research in Social and Administrative Pharmacy}\ }\textbf {\bibinfo {volume} {4}},\ \bibinfo {pages} {173} (\bibinfo {year} {2008})}\BibitemShut {NoStop}%
\bibitem [{\citenamefont {Caracelli}\ and\ \citenamefont {Greene}(1993)}]{Caracelli1993}%
  \BibitemOpen
  \bibfield  {author} {\bibinfo {author} {\bibfnamefont {V.~J.}\ \bibnamefont {Caracelli}}\ and\ \bibinfo {author} {\bibfnamefont {J.}~\bibnamefont {Greene}},\ }\bibfield  {title} {\bibinfo {title} {Data analysis strategies for mixed-method evaluation designs},\ }\href {https://doi.org/10.2307/1164421} {\bibfield  {journal} {\bibinfo  {journal} {Educational Evaluation and Policy Analysis}\ }\textbf {\bibinfo {volume} {15}},\ \bibinfo {pages} {195 } (\bibinfo {year} {1993})}\BibitemShut {NoStop}%
\bibitem [{\citenamefont {Onwuegbuzie}\ and\ \citenamefont {Combs}(2010)}]{Onwuegbuzie2010}%
  \BibitemOpen
  \bibfield  {author} {\bibinfo {author} {\bibfnamefont {A.}~\bibnamefont {Onwuegbuzie}}\ and\ \bibinfo {author} {\bibfnamefont {J.}~\bibnamefont {Combs}},\ }\bibfield  {title} {\bibinfo {title} {Emergent data analysis techniques in mixed methods research: A synthesis},\ }\href {https://doi.org/10.4135/9781506335193.n17} {\bibfield  {journal} {\bibinfo  {journal} {Handbook of Mixed Methods in Social and Behavioral Research}\ ,\ \bibinfo {pages} {397}} (\bibinfo {year} {2010})}\BibitemShut {NoStop}%
\bibitem [{\citenamefont {Wang}\ \emph {et~al.}(2008)\citenamefont {Wang}, \citenamefont {Zhang}, \citenamefont {McArdle},\ and\ \citenamefont {Salthouse}}]{wang2008investigating}%
  \BibitemOpen
  \bibfield  {author} {\bibinfo {author} {\bibfnamefont {L.}~\bibnamefont {Wang}}, \bibinfo {author} {\bibfnamefont {Z.}~\bibnamefont {Zhang}}, \bibinfo {author} {\bibfnamefont {J.~J.}\ \bibnamefont {McArdle}},\ and\ \bibinfo {author} {\bibfnamefont {T.~A.}\ \bibnamefont {Salthouse}},\ }\bibfield  {title} {\bibinfo {title} {Investigating ceiling effects in longitudinal data analysis},\ }\href {https://doi.org/10.1080/00273170802285941} {\bibfield  {journal} {\bibinfo  {journal} {Multivariate behavioral research}\ }\textbf {\bibinfo {volume} {43}},\ \bibinfo {pages} {476} (\bibinfo {year} {2008})}\BibitemShut {NoStop}%
\bibitem [{\citenamefont {Hiebert}\ \emph {et~al.}(2012)\citenamefont {Hiebert}, \citenamefont {Bezanson}, \citenamefont {O'Reilly}, \citenamefont {Hopkins}, \citenamefont {Magnusson},\ and\ \citenamefont {McCaffrey}}]{hiebert2012assessing}%
  \BibitemOpen
  \bibfield  {author} {\bibinfo {author} {\bibfnamefont {B.}~\bibnamefont {Hiebert}}, \bibinfo {author} {\bibfnamefont {L.}~\bibnamefont {Bezanson}}, \bibinfo {author} {\bibfnamefont {E.}~\bibnamefont {O'Reilly}}, \bibinfo {author} {\bibfnamefont {S.}~\bibnamefont {Hopkins}}, \bibinfo {author} {\bibfnamefont {K.}~\bibnamefont {Magnusson}},\ and\ \bibinfo {author} {\bibfnamefont {A.}~\bibnamefont {McCaffrey}},\ }\bibfield  {title} {\bibinfo {title} {Assessing the impact of labour market information: Final report on results of phase two (field tests)},\ }\href@noop {} {\bibfield  {journal} {\bibinfo  {journal} {Ottawa: Ressources humaines et d{\'e}veloppement des comp{\'e}tences Canada}\ } (\bibinfo {year} {2012})}\BibitemShut {NoStop}%
\bibitem [{\citenamefont {Kanevsky}(2016)}]{Kanevsky}%
  \BibitemOpen
  \bibfield  {author} {\bibinfo {author} {\bibfnamefont {L.}~\bibnamefont {Kanevsky}},\ }\href {https://www.sfu.ca/istld/faculty/resources/postpre.html} {\bibinfo {title} {Post-pre surveys}} (\bibinfo {year} {2016})\BibitemShut {NoStop}%
\bibitem [{\citenamefont {Barger}(2017)}]{barger2017reading}%
  \BibitemOpen
  \bibfield  {author} {\bibinfo {author} {\bibfnamefont {J.~M.}\ \bibnamefont {Barger}},\ }\bibfield  {title} {\bibinfo {title} {Reading is not essential to writing instruction},\ }\href@noop {} {\bibfield  {journal} {\bibinfo  {journal} {BAD IDEAS}\ ,\ \bibinfo {pages} {44}} (\bibinfo {year} {2017})}\BibitemShut {NoStop}%
\bibitem [{\citenamefont {Van~de Pol}\ \emph {et~al.}(2010)\citenamefont {Van~de Pol}, \citenamefont {Volman},\ and\ \citenamefont {Beishuizen}}]{van2010scaffolding}%
  \BibitemOpen
  \bibfield  {author} {\bibinfo {author} {\bibfnamefont {J.}~\bibnamefont {Van~de Pol}}, \bibinfo {author} {\bibfnamefont {M.}~\bibnamefont {Volman}},\ and\ \bibinfo {author} {\bibfnamefont {J.}~\bibnamefont {Beishuizen}},\ }\bibfield  {title} {\bibinfo {title} {Scaffolding in teacher--student interaction: A decade of research},\ }\href@noop {} {\bibfield  {journal} {\bibinfo  {journal} {Educational psychology review}\ }\textbf {\bibinfo {volume} {22}},\ \bibinfo {pages} {271} (\bibinfo {year} {2010})}\BibitemShut {NoStop}%
\bibitem [{\citenamefont {ARCHER}\ \emph {et~al.}(2015)\citenamefont {ARCHER}, \citenamefont {DEWITT},\ and\ \citenamefont {OSBORNE}}]{archer2015aspiration}%
  \BibitemOpen
  \bibfield  {author} {\bibinfo {author} {\bibfnamefont {L.}~\bibnamefont {ARCHER}}, \bibinfo {author} {\bibfnamefont {J.}~\bibnamefont {DEWITT}},\ and\ \bibinfo {author} {\bibfnamefont {J.}~\bibnamefont {OSBORNE}},\ }\bibfield  {title} {\bibinfo {title} {Is science for us? black students’ and parents’ views of science and science careers},\ }\href {https://doi.org/https://doi.org/10.1002/sce.21146} {\bibfield  {journal} {\bibinfo  {journal} {Science Education}\ }\textbf {\bibinfo {volume} {99}},\ \bibinfo {pages} {199} (\bibinfo {year} {2015})},\ \Eprint {https://arxiv.org/abs/https://onlinelibrary.wiley.com/doi/pdf/10.1002/sce.21146} {https://onlinelibrary.wiley.com/doi/pdf/10.1002/sce.21146} \BibitemShut {NoStop}%
\end{thebibliography}%

\appendix

\section{About Astrobites}\label{appendix:astrobites}

The Astrobites collaboration (or simply Astrobites) is a group of volunteer graduate students in astronomy and related fields who publish daily blog-style posts summarizing current astronomy research papers. Astrobites also publishes guides to specific topics and other resources relevant to career development and equity/inclusion. 
While the content from the Astrobites website is primarily targeted towards undergraduate astronomy majors, its audience has grown to include everyone from hobbyists to emeritus professors since \revv{its} founding in 2011. 

A central focus of Astrobites’ mission is to further efforts for inclusion and equity in astronomy. The organization aims to address the inaccessibility and technical language of research articles through summaries and ``guides" to different aspects of astronomy and academic professional development. Articles featured on Astrobites are also an excellent way for students to explore differences in audience between various forms of science writing, and for students -- both undergraduates reading the articles and graduate students serving as regular writers -- to practice writing skills. These skills are increasingly critical for becoming a successful scientist. Astrobites has also taken an active role in incorporating its pedagogical content, wide topical coverage, and scaffolded contextual resources into formal education \citep{sanders2012preparing,sanders2017incorporating}.

All content on Astrobites -- including the education resources used in this study -- is free and open access, publicly available on \href{astrobites.org}{astrobites.org}. Similar websites (known as ``Science Bites'' sites) are available in other disciplines.

\section{Full Assessment Text}\label{survey-app}

Questions not designated as ``interview'' or ``open-ended'' were answered on a Likert scale from 1 to 7, where 1 indicates strong disagreement, 4 is neutral, and 7 is strong agreement.

\underline{Perceived ability with jargon}
\begin{enumerate}
\item I am able to define technical terminology that appears in scientific literature.
\item Technical terms are a major barrier to my understanding of scientific literature. [Reversed]
\item I am able to use outside resources to help define technical terminology.
\item I am able to understand the technical details of a scientific study.
\item I find technical vocabulary in astronomy papers daunting. [Reversed]
\end{enumerate}

{\noindent \underline{Perceived ability with main takeaways}}
\begin{enumerate}
\setcounter{enumi}{5}
\item After reading a paper, I can summarize it well to a classmate in a science course.
\item I can understand the motivation behind a scientific paper.
\item I can understand how a given scientific paper contributes to the field at large.
\item I am able to figure out the main takeaways / key ideas of a scientific paper.
\item I am able to interpret the data on a plot and extract the key takeaway from it.
\item I know what the most important pieces of information are after reading a scientific research paper.
\end{enumerate}

{\noindent \underline{Perceived ability with conceptual understanding/intuition}}
\begin{enumerate}
\setcounter{enumi}{11}
\item I can see how concepts learned in class feature in astronomy papers.
\item I am able to understand astronomy concepts.
\item I am able to relate different astronomy concepts to each other.
\item I could evaluate whether the data on a plot seems reasonable given the scenario it represents.
\item I can interpret what the data on a plot is telling me about a concept.
\item I am able to draw analogies between concepts discussed in astronomy literature and concepts prevalent in everyday life.
\item I have a strong intuition for concepts in astronomy.
\end{enumerate}

{\noindent \underline{Perception of science communication ability}}
\begin{enumerate}
\setcounter{enumi}{18}
\item I feel comfortable discussing astronomy with my friends.
\item I feel comfortable discussing astronomy with my science classmates.
\item I feel comfortable discussing astronomy with my science instructors.
\item I could explain basic astronomy concepts to a non-scientist.
\item I can explain astronomy concepts to a classmate in a science course.
\item I can explain astronomy concepts to my science instructors.
\end{enumerate}

{\noindent \underline{Perceived ability within astronomy}}
\begin{enumerate}
\setcounter{enumi}{24}
\item I have what it takes to be an astronomer.
\item I am capable of succeeding in my astronomy courses.
\item I am able to use technical astronomy skills (tools, instruments, and/or techniques).
\item I am able to use scientific literature to guide my studies in astronomy.
\item I am capable of working on astronomy research.
\end{enumerate}

{\noindent \underline{Feelings of belonging within astronomy}}
\begin{enumerate}
\setcounter{enumi}{29}
 \item I see myself as someone who can contribute to science.
 \item I see myself as someone who can contribute to astronomy.
 \item I could see myself becoming an astronomer as my career.
 \item I feel a part of the group when I am with other astronomers.
 \item I have a strong sense of belonging to the community of astronomers.
 \item I feel like I belong in the field of astronomy.
 \end{enumerate}

{\noindent \underline{Open-Ended Questions}}
\begin{enumerate}
\setcounter{enumi}{35}
    \item How do you feel about reading scientific research papers?
    \item What comes to mind when you think of astronomy research?
    \item Do you think you are capable of being an astronomer (or related scientist)? Why or why not?
    \item Do you feel like you belong in astronomy? Why or why not?
    \item Did you enjoy this activity with Astrobites? Why or why not? [Post-Lesson Survey Only]
    \item What is your main takeaway from this activity? [Post-Lesson Survey Only]
    \item What did you find most useful about this activity? [Post-Lesson Survey Only]
    \item What was most confusing about this activity? [Post-Lesson Survey Only]
    \item How have your feelings about astronomy research changed as a result of this activity, if at all? [Post-Lesson Survey Only]
\end{enumerate}

{\noindent \underline{Instructor Interview Questions}}
\begin{itemize}
 \item Which Astrobites lesson plan did you implement, and did you make any changes to it?
  \item Overall, how do you think this lesson went in your classroom? 
 \item How do you think this activity / lesson plan impacted your students?
 \item Do you think your students are more prepared to engage with scientific research literature after this lesson? Why or why not? 
 \item Do you think your students enjoyed this activity with Astrobites? Why or why not?
 \item What did you find most useful about this lesson plan?
 \item What suggested changes do you have for this lesson plan?
\end{itemize}

\section{Further Information on Astrobites Assignments}\label{assignments-info}

Astrobites currently maintains four \rev{assignment templates} based on  prior published work \citep{sanders2017incorporating}. 
These \rev{templates} provide four pathways to engage students with material from Astrobites -- guided reading assignments, student research projects, a student writing assignment, and student presentations -- and are available in \href{https://docs.google.com/document/d/1jAr6VsjxYxOJFJfvTquvLF4id1x5\_JEvMM7eUPQnyNc/edit?usp=sharing} {this linked document}. For each \rev{assignment}, an overview of the learning objectives, an outline of how the plan can be adapted for different course levels, sample handouts for presenting the assignment to students, and a potential grading rubric \rev{are provided for instructors. }

The first \rev{assignment} (Periodic Reading Assignments) is designed to grow students' reading comprehension while strengthening their conceptual understanding of course material in the broader context of active research. Students build familiarity and confidence in their ability to monitor and keep pace with recent developments in the astronomy literature by reading a series of Astrobites articles and testing their understanding through directed questions. This \rev{assignment} can be easily adapted to different course levels by varying the frequency of assignments and altering the style of the associated questions (e.g., asking more directed questions for lower-division undergraduates compared to open-ended questions for graduate students).

In the second \rev{assignment} (Student Research Project), students select a topic and prepare a paper or presentation using articles from the Astrobites website to guide their research in a modified literature review. 
Students synthesize concepts and interpret figures and data from different articles for their report. 
Students build confidence in developing techniques to tackle new concepts
by using Astrobites articles to identify technical terminology (jargon) with which they are unfamiliar. Students then define jargon by creating a glossary. 
This \rev{assignment} is meant to teach students how to engage with full research articles in the future. 

For the third \rev{assignment} (Student Writing Assignment), students read one or more papers and write an Astrobites-like summary.
Much as with the second \rev{assignment}, students practice synthesizing information from multiple sources. However, they additionally develop valuable science communication and composition skills by compiling information into a bite-sized article.
Additionally, students actively practice reading the scientific literature directly. 
\revv{By breaking the assignment into smaller chunks} -- e.g., having students present drafts and outlines along the way -- instructors can help students settle into the reading and writing process more smoothly.

Finally, in the fourth \rev{assignment} (Student Presentation), students deliver an in-class presentation featuring a profile of an astronomer who has been interviewed by people from the Astrobites collaboration and a summary of a paper by the interviewee. 
While also incorporating the reading comprehension and communication skills that appear in the other \rev{assignments}, the focus of this \rev{assignment} is to support students' career preparation and develop a sense of belonging in the field. 
By sharing presentations describing the career trajectory and advice of successful astronomers hailing from diverse backgrounds, students are able to see that astronomy is for everyone, while also identifying and discussing important steps on the way to becoming a professional astronomer. For example, one study of the importance of role models for career preparation and aspiration of young Black British people is presented in \citet{archer2015aspiration} and was summarized in Astrobites in 2020. 

\section{Additional Response Quotes from Educators}\label{app:quotes}

Instructor comments on the utility of the \rev{assignment templates featuring accessible summaries}:
\begin{quote}
    ``I think it gave them something to chew on when I was doing lectures...Students were definitely engaged with the literature in a way that seemed positive.''
\end{quote}
\begin{quote}
    ``A student was like I just wanted to say that I really appreciated that there was a person there like working hard on accessibility and disabilities, and it's something that we don't see a lot of. And I was like, wow, these students are seeing themselves in accomplished astronomers. And that was cool for me as the instructor to witness.''
\end{quote}
\begin{quote}
     ``[Astrobites/accessible summaries] enabled the possibility to connect to the paper, because the paper was so inaccessible for the current know-how of the students."
\end{quote}
\begin{quote}
    ``I do [think they are more prepared] because they told me. But then from my perspective, I could see what they were writing asking deeper questions...I saw it [the improvement] in their work.''
\end{quote}
\begin{quote}
    ``I noticed when students would do presentations, they would reference the results from their classmates previous Astrobites presentations. And I feel like that is one indicator of being able to interact more strongly with the literature, is recalling the results from previous works and incorporating that into their own presentation.''
\end{quote}

\end{document}